\documentclass{JHEP}
\input epsf
\usepackage{epsfig}
\usepackage{amssymb}
\usepackage[active]{srcltx}

\newcommand{\bea}{\begin{eqnarray}}
\newcommand{\eea}{\end{eqnarray}}
\newcommand{\bean}{\begin{eqnarray*}}
\newcommand{\eean}{\end{eqnarray*}}

\def\O #1{\overline{#1}}

\def\D #1{\dot{#1}}

\def\det{\mathop{\rm det}}

\def\d{{\rm d}}

\def\th{{\theta}}
\def\bth{{\overline{\theta}}}
\def\a{{\alpha}}

\def\da{{\dot{\alpha}}}
\def\b{{\beta}}
\def\db{{\dot{\beta}}}
\def\c{{\gamma}}
\def\dc{{\dot{\gamma}}}

\def\d{\partial}
\def\rmd{{\rm d}}
\def\la{\lambda}
\def\eps{\epsilon}

\def\ald{{\dot\alpha}}
\def\bed{{\dot\beta}}

\def\sid{{\dot\rho}}

\preprint{
SNUST 031103\\
{\tt hep-th/0311275}}
\title{U(N) Instantons on ${\cal N}={1 \over 2}$ Superspace -- \\
exact solution \& geometry of moduli space \footnote{ This work
was supported in part by NSF grant PHY-0070928 (RB, BF, OL) and by
the KOSEF Leading Scientist Grant (SJR).}}
\author{Ruth Britto ${}^a$, \, Bo Feng ${}^a$, \, Oleg Lunin ${}^a$, \,
Soo-Jong Rey ${}^{a,b}$\\
~~~~~~~~~~~~~~\\
${}^a$ Institute for Advanced Study\\
 Einstein Drive, Princeton NJ 08540  USA\\
~~~~~~~\\
${}^b$ School of Physics \& BK-21 Physics Division\\
 Seoul National University, Seoul 151-747 KOREA\\
 ~~~~~~\\
 \email{britto, fengb, lunin@ias.edu \hskip0.5cm sjrey@gravity.snu.ac.kr} }
\abstract{We construct the exact solution of one (anti)instanton
in ${\cal N}={1 \over 2}$ super Yang-Mills theory defined on
non(anti)commutative superspace. We first identify ${\cal N}={1
\over 2}$ superconformal invariance as maximal spacetime symmetry.
For gauge group U(2), the SU(2) part of the solution is given by
the standard (anti)instanton, but the U(1) field strength also
turns out nonzero. The solution is SO(4) rotationally symmetric.
For gauge group U(N), in contrast to the U(2) case, we show that
the entire U(N) part of the solution is deformed by
non(anti)commutativity and fermion zero-modes. The solution is no
longer rotationally symmetric; it is polarized into an axially
symmetric configuration because of the underlying
non(anti)commutativity. We compute the `information metric' of one
(anti)instanton. We find that the moduli space geometry is
deformed from hyperbolic space $\mathbb{H}_5$ (Euclidean anti-de
Sitter space) in a way anticipated from reduced spacetime
symmetry. Remarkably, the volume measure of the moduli space turns
out to be independent of the non(anti)commutativity. Implications
for D-branes in Ramond-Ramond flux background and the
gauge-gravity correspondence are discussed.}

\begin{document}

\section{Introduction}
Recently, there has been considerable development in understanding
superstrings and D-branes in the background of Ramond-Ramond flux.
Take Type IIB superstring theory compactified on $X \times
\mathbb{R}^4$, where $X$ is a Calabi-Yau threefold. Turn on a
Ramond-Ramond 5-form $G^+_5$ on a holomorphic cycle of $X$; the flux
corresponds on $\mathbb{R}^4$ to a self-dual graviphoton flux.
Introduce D3-branes whose worldvolume fills $\mathbb{R}^4$. For
closed strings, the graviphoton flux deforms the four-dimensional
${\cal N}=2$ supersymmetry algebra, in which half of the
supersymmetry is realized nonlinearly. For open strings on the
Euclidean D3-branes, the graviphoton flux deforms the ${\cal N}=1$
supersymmetry \cite{dijkgraafvafa}--\cite{berkovitsseiberg}.
The deformation induces
non(anti)commutativity among the Grassmann-odd coordinates,
\bea \{ \th^\a, \th^\b \} = C^{\a\b}, \qquad \{\bth^\da, \bth^\db
\} = 0, \qquad \{\th^\a, \bth^\da \} = 0, \label{nac} \eea
and breaks the underlying ${\cal N}=1$ supersymmetry to ${\cal
N}={1 \over 2}$. Accordingly, the low-energy worldvolume dynamics
of Euclidean D3-brane is governed by a non(anti)commutative super
Yang-Mills theory with ${\cal N}={1 \over 2}$
supersymmetry.\footnote{For recent works dealing with various
aspects of theories with ${\cal N}={1 \over 2}$ supersymmetry, see
\cite{therest}.} The ${\cal N}={1 \over 2}$ super Yang-Mills
theory is then defined by\footnote{Our conventions and notation
are collected in Appendix A.} the action functional \cite{seiberg}
\bea S_{\rm YM} = -\int_{\mathbb{R}^4} {\rm Tr} \left[ {i \tau
\over 8 \pi} W^\a \star W_\a \right]_{\th^2} + {\rm Tr} \left[{i
\overline{\tau} \over 8 \pi} \overline{W}_\da \star
\overline{W}_\da \right]_{\bth^2}, \label{lagformal} \eea
where the non--(anti)commutativity (\ref{nac}) is realized in terms
of the $\star$-product:
\bea A(\th) \star B(\th) \equiv A(\th) \exp \left( -{1 \over 2}
C^{\a\b} \overleftarrow{\partial \over \partial \th^\a}
\overrightarrow{\partial \over \partial \th^\b} \right) B(\th).
\label{star} \eea
Though the non--(anti)commutativity parameter $C^{\a\b}$ carries a
nonzero scaling dimension, it turns out that, to all orders in
perturbation theory, the non--(anti)commutative deformation of a
renormalizable ${\cal N}=1$ supersymmetric field theory remains
renormalizable \cite{renorm,berey}. Intuitively, in Wilson's
renormalization-group
viewpoint, the renormalizability is explainable by chirally
asymmetric assignments of scaling dimensions, a possibility made
available by the non--(anti)commutative deformation \cite{berey}.

We are primarily interested in the low-energy dynamics of ${\cal
N}={1 \over 2}$ supersymmetric gauge theory. The motivation comes
largely from two sides. First, the dynamics by itself is quite
interesting and may provide a novel way of interpolating between
gauge dynamics with ${\cal N}=1$ and ${\cal N}=0$ supersymmetries.
Second, the dynamics may probe the Calabi-Yau geometry with the
Ramond-Ramond flux $G_5$ turned on. It then becomes imperative
to understand instantons in ${\cal N}={1\over 2}$ supersymmetric
gauge theories.

In exploring instantons in ${\cal N}={1 \over 2}$ super Yang-Mills
theory, a variety of interesting questions arise. At the
ultraviolet fixed point, the ${\cal N}=1$ theory is known to
promote the Poincar\'e supersymmetry to superconformal
symmetry. The superconformal symmetry algebra is SU$(4|1)$ and
involves 16 bosonic generators and 8 fermionic ones. We will show
that, once the non--(anti)commutativity is turned on, the symmetry
algebra is reduced to an ${\cal N}={1 \over 2}$ superconformal
algebra. In this reduced symmetry algebra, the special
conformal and the chiral SU$(2)_L$ generators (as well as their
fermionic partners) are removed, and the dilatation and the
R-symmetry generators combine into a single generator of the form
dictated precisely by the new scaling dimension assignment put
forward in \cite{berey}.

Despite being deformed by the non--(anti)commutativity, the
instanton carries integrally quantized topological charge,
\bea {\cal Q}_{\rm instanton} =  \int_{\mathbb{R}^4} {1 \over 8
\pi^2} {\cal F} = \mathbb{Z}, \qquad {\rm where} \qquad {\cal F}
:= {\rm Tr}_{\rm U(N)} F \wedge F \, . \label{topcharge} \eea
This is in full accord with the Atiyah-Singer index theorem and
assures that the deformed anti-instantons are analytic in
$C^{\a\b}$. There is a good rationale behind this. The instanton
supports fermionic zero-modes. What is nontrivial in the present
context is that the instanton solution is corrected by the
fermionic zero-modes. Accordingly, the topological charge {\sl
density} ${\cal F}$ itself depends not only on the bosonic
zero-modes but also on (even powers of) fermionic zero-modes.
Moreover, since the non(anti)commutative superspace is not
invariant under the full SO(4)$=$SU$(2)_L \times$SU$(2)_R$
rotation group, the instanton would not be rotationally symmetric
in general.

It turns out that the above two issues are intimately related. For
the gauge group $G=$U(2), we will find that the one-instanton
solution exhibits trivial dependence on the non(anti)commutativity
--- the SU(2) part of the solution is the standard instanton, and
the U(1) part is a multipole configuration induced through
fermionic zero-modes and non(anti)commutativity. The U(1) part cannot
contribute to the topological charge; this is how the deformed
instanton remains consistent with the Atiyah-Singer index
theorem. The entire configuration is spherically symmetric, viz.
the U(2) instanton exhibits accidentally larger spacetime
symmetries.

For a gauge group of higher rank, $G=$U($N\!\ge\!3$), the story is
far more interesting and intricate. Start with the standard SU(2)
instanton embedded in U(N), and examine how the
non(anti)commutativity deforms the instanton configuration. In
stark contrast to the $G=$U(2) case, we find that the
one-instanton solution is deformed not only in the U($N-2$) part
but in the SU(2) part as well! As the attentive reader will
notice, this leads immediately to the possibility that the
topological charge density, and hence the charge itself, depends
on the fermionic zero-modes. We shall find that the topological
charge density indeed depends on the fermionic zero-modes, but the
charge itself is actually independent of them. The way this is
made possible turns out to be nicely intertwined with the absence
of rotational invariance in the problem. We will demonstrate that
the deformation induced by the non(anti)commutativity polarizes
the topological charge density into a sort of dipole
configuration. The deformation is axially symmetric  but is fully
compatible with the antichiral SU$(2)_R$ invariance. Thus, once
integrated over $\mathbb{R}^4$, the dipolar deformation is washed
out, retaining only a spherically symmetric contribution from the
standard SU(2) instanton. The latter yields integrally quantized
topological charge.

One can learn more physics from the topological charge {\sl
density} ${\cal F}$, since it is a function of bosonic and
fermionic zero-modes in addition to being a function of
coordinates on $\mathbb{R}^4$. What one expects to be modified by
the non(anti)commutative deformation is the geometry of the
one-instanton moduli space. To explore the issue, we compute the
information metric of one instanton, first put forward by Hitchin
\cite{Hitchin}. For small instantons, we find that the information
metric approaches that of a 5-dimensional hyperbolic space
$\mathbb{H}_5$ (Euclidean anti-de Sitter space, $\mbox{AdS}_5$).
The asymptotic isometry SO(5,1) is much bigger than the ${\cal
N}={1 \over 2}$ superconformal symmetry, so one expects the
interior of the moduli space not to retain the $\mathbb{H}_5$
geometry globally. Indeed, we find that the geometry of the moduli
space is deformed for larger instantons --- by
non(anti)commutativity, not only each metric component is
deformed, but also off-diagonal components of the metric are
induced. In fact, these corrections are fully compatible with the
symmetries that underlie the theory: ${\cal N}={1 \over 2}$
supersymmetry, R-(pseudo)symmetry and dilatation symmetry.
Remarkably, after a suitable change of zero-mode variables, the
volume measure on the moduli space turns out to be {\sl
independent} of the non(anti)commutative deformation! This
observation bears implications for Maldacena's gauge-gravity
correspondence,  on which we will elaborate in section 7.

We have organized the present paper as follows. In section 2, we
analyze the spacetime symmetry for theories defined on ${\cal
N}={1 \over 2}$ superspace. We find that the underlying ${\cal
N}=1$ superconformal symmetry is broken explicitly to `half' of
it, yielding what we call ${\cal N}={1 \over 2}$ superconformal
symmetry. This symmetry will provide a useful guideline for
constructing ${\cal N}={1 \over 2}$ instantons in subsequent
sections. In section 3, we derive self-duality and
anti-self-duality equations by localizing the action on
appropriate supersymmetric loci in field configuration space. In
section 4, we construct the instanton for the gauge group
$G=$U(2). This is a special situation where the instanton calculus
becomes almost trivial, due in major part to the trivial
back-reaction of the fermion quasi-zero-modes to the undeformed
instanton. In section 5, we construct the instanton for gauge
groups of higher rank, namely $G=$U($N$) for $N \ge 3$. To
illustrate the general strategy, we first set superconformal
fermionic zero-modes to zero, and consider perturbations by
supersymmetry fermionic zero-modes only. In section 6, we include
the superconformal fermionic zero-modes and find the exact
instanton solution for gauge group $G=$U($N$). In both sections,
we set out analytic strategy in a way adaptable for the
Atiyah-Drinfeld-Hitchin-Manin (ADHM) method \cite{ADHM},
relegating a direct ADHM construction for multi-instantons to
future work. In section 5, we present the exact solution for the
${\cal N}={1 \over 2}$ anti-instanton for the gauge group $G=$
U($N$). In section 7, we study the profile on $\mathbb{R}^4$ of
the topological charge density. We find that the density exhibits
dipolar polarization, whose size is set by the
non(anti)commutative deformation and whose symmetry fits precisely
with the underlying spacetime symmetries. We next study the
density profile on the instanton moduli space by computing
Hitchin's information metric. We find that the geometry of the
moduli space asymptotes to that of $\mathbb{H}_5$ (Euclidean
anti-de Sitter space) near the boundary. In the interior, the
geometry is deformed by the non(anti)commutativity, but again in a
form fully compatible with the underlying spacetime symmetries. We
discuss aspects of this observation in the context of Maldacena's
gauge-gravity correspondence. In the appendices, we collect
conventions and notation, undeformed SU(2) instanton and
anti-instanton solutions, and some essential steps of the
computation for obtaining the exact U(N) solution presented in
section 6.

During the progress of this work, a paper by Imaanpur
\cite{imaanpur} appeared, overlapping with part of our section 4.
We find agreement (modulo errors and numerical factors) wherever
both results overlap. Also, while this work was being written up,
a paper by Grassi et.\ al.\ \cite{grassi} appeared, again
overlapping with part of our section 4. We believe that our
motivation, results and interpretation are in strong contrast to
theirs.
\section{${\cal N}={1 \over 2}$ Superconformal Algebra}

We begin with observations regarding symmetry associated to the
non--(anti)commutative ${\cal N}={1 \over 2}$ superspace. The
underlying (anti)commutative ${\cal N}=1$ superspace is
parametrized by the coordinates $(x^{\a\da}, \th^\a, \bth^\da)$ ---
bosonic, chiral and antichiral fermionic coordinates. The
superspace displays ${\cal N}=1$ Poincar\'e supersymmetry. If
dilatation invariance is additionally endowed, the symmetry is
enlarged to ${\cal N}=1$ superconformal symmetry. This is the
symmetry we are most interested in. For example, if a theory
defined on the superspace has no mass scale, classically and/or
quantum-mechanically, then the operators and states of the theory are
organized in irreducible representations of the superconformal
group SU$(2,2|1)$ or SU$(4|1)$.

Once the non--(anti)commutativity deformation is turned on for the
chiral fermionic coordinates as in (\ref{nac}), the ${\cal N}=1$
supersymmetry is broken to ${\cal N}={1 \over 2}$ supersymmetry.
This is seen by examining the deformation of the ${\cal N}=1$
supersymmetry algebra. Though the algebra among the ${\cal N}=1$
superspace derivatives
\bea D_\a = +\partial_\a + 2i \bth^\da
\partial_{\a\da}, \qquad \overline{D}_\da = - \overline{\partial}_\da
\nonumber \eea
remains unaffected:
\bea \left\{ D_\a, D_\b \right\}_\star &=& 0 \nonumber \\
\left\{ \overline{D}_\da, \overline{D}_\db \right\}_\star &=& 0
\nonumber \\
\left\{ D_\a, \overline{D}_\da \right\}_\star &=& - 2 P_{\a \da},
\nonumber \eea
the algebra among the ${\cal N}=1$ supersymmetry charges
\bea Q_\a = +\partial_\a, \qquad \overline{Q}_\da = -
\overline{\partial}_\da + 2 i \th^\a \partial_{\a \da} \eea
now obey deformed anticommutation relations:
\bea \left\{ Q_\a, Q_\b \right\}_\star &=& 0 \nonumber \\
\left\{ Q_\a, \overline{Q}_\ald \right\}_\star &=& 2 P_{\a\ald} \nonumber \\
\left\{ \overline{Q}_\ald, \overline{Q}_\bed \right\}_\star &=&
4 C^{\a\b} P_{\a\ald} P_{\b\bed} \, .  \label{brokenn=1}\eea
The last relation indicates that repeated action of the
$\overline{Q}$ supercharges is ill-defined, violating the Leibnitz
rule.\footnote{Note, however, that a single action
of the $\overline{Q}$ charge is meaningful. In particular, the
second relation in (\ref{brokenn=1}) indicates that acting
$\overline{Q}_\da$ on the ${\cal N}={1 \over 2}$ supercharges
$Q_\a$ generates translation on $\mathbb{R}^4$. In the next
section, we will use this observation to derive
instanton equations.} As such, (\ref{brokenn=1}) does not form an algebra. The
subalgebra generated by the $Q_\a$'s is still preserved, and this
defines precisely the chiral ${\cal N}={1 \over 2}$ supersymmetry
algebra.

Implicit in the above route to the ${\cal N}={1 \over 2}$
supersymmetry is that the non--(anti)commutative superspace is
parametrized in terms of so-called chiral coordinates $(y, \th)$,
where $y, \overline{y}$ refer to chiral and antichiral
Grassmann-even coordinates:
\bea y^{\a\da} := (x^{\a\da} -2i  \th^\a \bth^\da) \qquad
\mbox{and} \qquad \overline{y}^{\a\da} := (x^{\a\da} +2i \th^\a
\bth^\da). \nonumber \eea
Various considerations point to this as the correct choice. First,
in terms of the chiral coordinates, as observed by Seiberg
\cite{seiberg}, chiral and antichiral superfields are
definable in a manner compatible with the non--(anti)commutative
$\star$-product (\ref{star}). Second, the ${\cal N}={1 \over 2}$
superspace can be parametrized uniquely by $(y, \th^\a)$, for
which the ${\cal N}={1 \over 2}$ supersymmetry acts as a chiral
Grassmann-odd translation:
\bea  (y^{\a\da}, \th^\a) \longrightarrow (y^{\a\da}, \th^\a +
\varepsilon^\a). \nonumber \eea

Having identified the canonical choice of coordinates on ${\cal
N}={1 \over 2}$ superspace, we are now ready to analyze spacetime
symmetries. In doing so, we will come across the idea
\cite{berey} behind the intuitive proof of renormalizability of
non--(anti)commutative field theories. In \cite{berey}, it was
argued that the most natural assignment of scaling dimensions is
such that $\th^\a$ is dimensionless, and hence $C^{\a\b}$ also is
dimensionless. The new scaling dimension is now measured as a
particular linear combination of the conventional scaling
dimension and the R-symmetry charge. In other words, the new
dilatation operator $D_{\rm new}$ is a linear combination of the
conventional dilatation operator $D$ and the R-symmetry charge
$R$. We will now show that this is precisely what comes out of
the analysis of spacetime symmetries associated with ${\cal N}={1
\over 2}$ superspace.

We claim that, on the non--(anti)commutative superspace, the
spacetime symmetry is realized on the following set of generators,
\bea \overline{M}_{\da\db}, \quad D_{\rm new} \equiv D - {1 \over
2} R, \quad P_{\a\da}, \quad Q_\a, \quad \overline{S}^\da,
\label{n1/2sc} \eea
which we refer to as the ${\cal N}={1 \over 2}$ superconformal
symmetry generators. Notice that the special conformal
transformation is no longer part of the symmetry, so the symmetry
group does not encompass the conformal transformations. Rather, it
should be viewed as supersymmetrization of the dilatation
transformation. This implies that, at a renormalization-group
fixed point, scale invariance of non--(anti)commutative field
theories would not be enhanced to superconformal invariance, in
stark contrast to the more familiar quantum field theories
\cite{polchinski}. Implicit to the latter is the requirement of
unitarity and Poincar\'e invariance, but these are precisely what
we drop in non--(anti)commutative field theories.

The proof of (\ref{n1/2sc}) is elementary. Begin by realizing
the ${\cal N}=1$ superconformal generators, again in the basis
of chiral superspace coordinates $(y, \th, \bth)$. They are
\bean \overline{M}_{\da\db} &=&
\frac{1}{2}{y^{\c}}_{(\da}\d_{\c\db)} - \bth_{(\da} \bar\d_{\db)}
; \hskip0.8cm P_{\a\da} = i \d_{\a\da} ; \hskip0.8cm {M}_{\a\b} =
\frac{1}{2}{y_{(\a}}^{\dc} \d_{\b)\dc} -
\th_{(\a} \d_{\b)}\\
K_{\a\da} &=& -i {y_\a}^{\db} {y^\b}_{\da} \d_{\b\db}
 + 2i y^\b_{~\da}\th_\b\d_\a + 2iy_{\a\da}(\th^\b\d_\b + \bth^\db\bar\d_\db)
  + 4 \th_\a \bth^2 \bar\d_\da + 2iy_\a^{~\db}\bth_\db\bar\d_\da \\
R \,\, &=& i\th^\a\d_\a - i\bth^\da\bar\d_\da  ; \hskip3.5cm  D
\,\, = -\frac{i}{2}y^{\a\da}\d_{\a\da} + \frac{i}{2}\th^\a\d_\a
+ \frac{i}{2}\bth^\da\bar\d_\da  \\
Q_\a &=& \d_\a; \hskip5.5cm \overline Q_\da = -\bar\d_\da+2i\th^\a \d_{\a\da}  \\
\overline{ S}_\da &=& y^\a_{~\da}Q_\a+2i\bth^2 \overline{D}_\da;
\hskip3.1cm S_\a = -(y_\a^{~\db}+4i\th_\a\bth^\db)\bar{Q}_\db +
2i\th^2 D_\a \, .  \nonumber \eean
It is now straightforward to check $\star$-(anti)commutators among
these generators. In doing so, we need to take into account the
non--(anti)commutativity among the $\th^\a$'s as in (\ref{nac}).
As mentioned above, all other coordinates (anti)commute {\sl
provided} one adopts the chiral superspace coordinates. One finds
by straightforward computation that the algebra closes on the
subset (\ref{n1/2sc}), whose non-vanishing $\star$-commutators are
\bea
\left[\overline{M}_{\da\db},\overline{M}_{\dc\sid}\right]_\star
&=& \eps_{\da(\dc}\overline{M}_{\sid )\db}
+ \eps_{\db (\dc}\overline{M}_{\sid) \da}, \nonumber \\
\left[ \overline{M}_{\da\db}, \, P_{\c\dc}  \right]_\star &=&
4 P_{\c(\da} \eps_{\db)\dc}, \nonumber\\
\left[\overline{M}_{\da\db},\, \overline{ S}_\dc \, \right]_\star &=&
\eps_{(\da \dc} \overline{ S}_{\db)}, \nonumber\\
\left[\, P_{\a\da} \, , \, \overline{ S}_\db \, \right]_\star &=&
2i \eps_{\da\db} Q_\a, \nonumber\\
\left[D_{\rm new}, P_{\a\da}\right]_\star &=& -i P_{\a\da},\nonumber\\
\left[D_{\rm new },\, \overline{ S}_\da \, \right]_\star &=& +i
\overline{S}_\da, \nonumber \eea
while the rest do not even form an algebra because the deformation
induces terms violating the Leibniz rule, much as the last
relation in (\ref{brokenn=1}).
We notice that only those generators whose expressions do not contain
the coordinate $\theta$ are the ones preserved by the deformation.

The algebra (\ref{n1/2sc}) shows that translational invariance is
retained, while (half of) SO(4) rotational and special conformal
invariance are lost. Therefore, one expects that an instanton in
${\cal N}={1 \over 2}$ would produce only those zero-modes
associated with these generators, and span the coordinates of
one-instanton moduli space. In the following sections, we shall
see how this restricted symmetry plays out in adding deformation
terms to the one-instanton solution and the metric on the moduli
space.

\section{(Un)deformed Instanton Equations}
In this section, we set up the problem of constructing instantons
and anti-instantons in ${\cal N}={1 \over 2}$ super Yang-Mills
theory. First, and to help set up our notation, we recapitulate
the definition of the theory. We then derive instanton and
anti-instanton equations and argue that with the self-dual
deformation by $C^{\a\b}$, the anti-self-duality equations are
deformed, while the self-duality equations are not.

Expanding in terms of the component fields, the action functional
of the non--(anti)commutative Yang-Mills theory (\ref{lagformal}) is
given by \cite{seiberg}:
\bea S_{\rm YM} &=& {{\rm Im}\, \tau \over 4 \pi}
\int_{\mathbb{R}^4} {\rm Tr} \! \left[ - {1 \over 2} F_{mn} F_{mn}
-i\overline{\lambda \lambda} \, C_{mn} F_{mn} + \frac{1}{4}
(\overline{\lambda \lambda})^2 \, C_{mn} C_{mn}
-i{\bar\la}{\bar\sigma}^m\nabla_m\la+ D^2 \right]
\nonumber\\
&-& i{{\rm Re}\, \tau  \over 8 \pi} \int_{\mathbb{R}^4} {\rm Tr}
\, F_{mn} \widetilde{F}_{mn}. \label{lagcomp} \eea
Here, we take the gauge group to be $G=$U($N$).\footnote{Under the
$\star$-product (\ref{star}), the enveloping algebra involving the
Lie algebra $su(N)$ is $u(N)$. We adopt the conventions that the
$u(N)$ generators ${\bf T}^a$ ($a=0,1, \cdots, N^2-1)$ are
normalized as ${\rm Tr} {\bf T}^a {\bf T}^b = {1 \over 2}
\delta^{ab}$, and the gauge covariant derivatives are
$\nabla_{\a\da} =
\partial_{\a\da} + \frac{i}{2}[A_{\a\da}, \cdot ]$.} We also denote
the coupling parameters in the convention of Minkowski spacetime
\bea {\rm Re} \, \tau \equiv {1 \over 2}( \tau + \overline{\tau}) ,
\qquad {\rm Im}\,  \tau \equiv {1 \over 2 i} (\tau -
\overline{\tau}), \eea
but, because the theory is defined on Euclidean space
$\mathbb{R}^4$, we interpret them as referring to {\sl two}
independent complex coupling constants $\tau, \overline{\tau}$. In
particular, by taking $\tau$ or $\overline{\tau}$ to infinity, one
can localize the super Yang-Mills action functional to $D_{(\a}
W_{\b)} = 0$ or $\overline{D}_{(\da} \overline{W}_{\db)} = 0$
field configurations, viz.  anti-self-duality and self-duality
configurations. Decompose the gauge field strength into self-dual
and anti-self-dual parts:
\bea F^{(+)}_{mn} &\equiv& {1 \over 2} \left( F +^* F \right)_{mn}
= {1 \over 2} F_{\a\b} \sigma^{\a\b}_{mn} \nonumber
\\
F^{(-)}_{mn} &\equiv& {1 \over 2} \left( F - ^*F \right)_{mn} = {1
\over 2} F_{\da\db} \overline{\sigma}^{\da\db}_{mn}. \nonumber
\eea
We will now derive the localization to self-duality or
anti-self-duality configurations explicitly.

\subsection{Antiholomorphic instanton from anti-self-duality}
To derive the anti-self-dual equations, we arrange the action
functional (\ref{lagcomp}) into perfect squares involving
$F^{(+)}$ as \footnote{This method was also considered
independently in \cite{imaanpur}.}
\bea\label{ConvenLagr} S_{\rm YM} = {{\rm Im}\, \tau \over 4 \pi}
\int_{\mathbb{R}^4} {\rm Tr} \left[ - \left(
F^{(+)}_{mn}+\frac{i}{2}C_{mn}\, \overline{\lambda \lambda}
\right)^2 - i\la \, \sigma^m \nabla_m {\bar\la} + D^2\right] -
{i\overline{\tau} \over 4 \pi} \int_{\mathbb{R}^4} {\rm Tr} F
\wedge F. \eea
The last term is a topological invariant, so the action functional
has a critical point at which
\bea F^{(+)}_{mn}+\frac{i}{2}C_{mn} \, \overline{\lambda \lambda}
= 0, \qquad  \sigma^m\nabla_m\overline{\la} = 0, \qquad \la = 0,
\qquad D=0. \label{asdeqn} \eea
These equations define anti-self-duality conditions, whose
solutions are anti-instantons. Notice that, compared to the ${\cal
N}=1$ supersymmetric anti-self-dual equations, (\ref{asdeqn}) are
deformed by the terms proportional to the self-dual
non--(anti)commutativity parameter $C_{mn}$. Notice in
(\ref{ConvenLagr}) that, though expressed into a perfect square,
the first term is not positive definite -- the
non(anti)commutativity parameter $C_{mn}$ is in general
complex-valued and the gaugino $\overline{\lambda}$ is no longer a
Majorana fermion in Euclidean space. So, the critical point
(\ref{asdeqn}) should be understood as enhanced symmetry point
rather than a minimum action configuration. Closely related to
this, in the first work of \cite{renorm}, it was shown that the
supersymmetry state is not a configuration of minimum energy but
of enhanced symmetry. In fact, owing to the
non(anti)commutativity, the energy (defined as eigenvalues of the
Hamiltonian) is in general complex-valued.

The anti-self-dual equations are also derivable by considering the
${\cal N}={1 \over 2}$ supersymmetry transformations. In the action-functional
(\ref{lagformal}), the chiral field strength superfield is given
in the Wess-Zumino gauge as
\bea W_\a (y, \th) = - i \lambda_\a(y) + \left[\th_\a D(y) - i
\left(F_{\a\b}(y) + {i \over 2} C_{\a\b}
\overline{\lambda\lambda}(y) \right)\th^\b \right] + \th\th
\nabla_{\a\da}\overline{\lambda}^\da(y) \, . \nonumber \eea
Component fields transform under the ${\cal N}={1 \over 2}$
supersymmetry as
\bea \delta \la^\a &=&i \eps^\a D
+2\Big(F^{\a\b}+\frac{i}{2}C^{\a\b} \overline{\lambda \lambda} \Big) \eps_\b\nonumber\\
\delta F_{\a\b} \! &=& - i \eps_{(\a} \nabla_{\b)\bed}
\overline{\la}^\bed \nonumber \\
\delta D \,\,\, &=&-\eps^\a \nabla_{\a\bed}
{\overline\la}^\bed \nonumber\\
 \delta F_{\da\db} &=& 0 \nonumber \\
\delta  \overline{\la}^\da \,\, &=& 0. \label{1/2susytransf} \eea
Take in (\ref{lagformal}) the limit ${\tau} \rightarrow \infty$.
In this limit, field configurations are localized to
\bea 0 &=&  {\rm Tr} \epsilon^{\a\b}\left[ W_\a \star W_\b
\right]_{\th^2}
\nonumber \\
&=& {\rm Tr} \epsilon^{\a\b}\left[ -{1 \over 2} (\delta_\a
\lambda)(\delta_\b \lambda)  -i \la_\a \,
\delta_\b D \right].
\nonumber \eea
We find that the configuration is localized where the ${\cal N}={1
\over 2}$ supersymmetry variations vanish. Moreover, inferring the
supersymmetry transformation rules (\ref{1/2susytransf}), the
localization locus is precisely the critical point specified by
(\ref{asdeqn}). We will refer to a configuration satisfying the
anti-self-duality conditions (\ref{asdeqn}) as an {\sl
antiholomorphic instanton}, since its strength is proportional to
multiple powers of $\exp(- 2 \pi i \overline{\tau})$.

Notice that each equation in (\ref{asdeqn}) is preserved under the
${\cal N}={1 \over 2}$ supersymmetry transformations
(\ref{1/2susytransf}), but that does not mean that the functional
form of the solution is preserved too. In fact, we shall find in
the next section that the solution is corrected through the
$C^{\a\b}$-dependent fermion bilinear term in (\ref{asdeqn}). This
correction has the following implications. Suppose we start with
the ordinary instanton solving the anti-self-duality equation
$F^{(+)} = 0$. This instanton is an $L^2$-normalizable solution of
the $\overline{\la}$ equation in (\ref{asdeqn}). As is evident
from (\ref{asdeqn}), this solution does not break the ${\cal N}={1
\over 2}$ supersymmetry; in particular, $\delta F_{\a\b} = \delta
D = 0$. It is illuminating to recast this from the underlying
${\cal N}=1$ supersymmetry viewpoint. The $L^2$-normalizable
$\overline{\la}$ zero-mode solution breaks `spontaneously' the
antichiral supersymmetry (generated by $\overline{Q}_\da$), but
this is already broken `explicitly' as the
non--(anti)commutativity deformation is turned on. As such, we
will refer to the $L^2$-normalizable $\overline{\la}$ solution
solving (\ref{asdeqn}) as {\sl quasi zero-modes}. As discussed in
the previous paragraph, the ${\cal N}={1 \over 2}$ supersymmetry
does not preclude back-reaction of these quasi zero-modes to the
first equation in (\ref{asdeqn}). It then modifies the vector
potential one started with. Analogously, there will be quasi
superconformal zero-modes, which will also react back to the
bosonic equations in (\ref{asdeqn}).
\subsection{Holomorphic instanton from self-duality}
To derive the self-duality conditions, we arrange the action
functional (\ref{lagcomp}) into terms involving $F^{(-)}$ as
\bea S_{\rm YM} = {{\rm Im}\, \tau\over 4 \pi} \int_{\mathbb{R}^4}
{\rm Tr} \left[ -\left( F^{(-)}_{mn} \right)^2 +
\overline{\lambda}^\da [M, \overline{\lambda}_\da \} -
i\overline{\la}^\da \, \overline{\sigma}^m_{\da\a} \nabla_m \la^a
+ D^2\right] + {i{\tau}\over 4 \pi} \int_{\mathbb{R}^4} {\rm Tr}
\, F \wedge {} F \, ,  \nonumber \eea
where the kernel $M$, which is an (anti)commutator depending on
$A_m, \overline{\lambda}$, is defined by
\bea [M, \, \cdot \} := -{1 \over 2} C^{mn} \{F_{mn}, \, \cdot \}
+ C^{mn} \{A_m, \nabla_n \, \cdot \} + {i \over 4}C^{mn} \{ A_m, [ A_n,
\, \cdot ]\} - {i\over 16} C_{mn}^2 \{ \overline{\la \la}, \,
\cdot \}. \nonumber \eea
Again, the last term is a topological invariant, so the action
functional has a critical point at which
\bea F^{(-)}_{mn} = 0, \qquad i \overline{\sigma}^m \nabla_m \la =
0, \qquad \overline{\la} = 0, \qquad D = 0. \label{sdeqns} \eea
These equations are the standard self-duality equations, and being
independent of $C^{\a\b}$, they are apparently unmodified by the
non--(anti)commutativity deformation.

Actually, the self-duality equations (\ref{sdeqns}) involve some
highly nontrivial effects arising from the non--(anti)commutative
deformation. This can be seen by resorting to the `broken'
antichiral supersymmetry generated by $\overline{Q}$. The
antichiral field strength superfield is given in the Wess-Zumino
gauge by
\bea \overline{W}_\da = i \overline{\la}_\da(\O y) + \left[
\bth^\da D(\O y) - i F_{\da\db}(\O y) \bth^\db \right] + \bth^2 [M,
 \overline{\la}_\da \} \, . \nonumber \eea
Under the antichiral supersymmetry (generated by $\overline{Q}$),
the component fields transform as
\bea \overline\delta\bar\lambda^\da &=& -i\overline\eps^\da D -
2 F^{\da\db}\overline\eps_\db \nonumber \\
\overline\delta A_{\a\da} &=& -2i \lambda_\a \overline\eps_\da \nonumber \\
\overline\delta F_{\da\db} &=& i \overline\eps_{(\da}
\nabla_{\a\db)}\lambda^\a \nonumber \\
\overline\delta D \, \, &=& -\overline\eps^\da \nabla_{\a\da}
\lambda^\a + i \bar\eps^\da \left[M,\bar\lambda_\da \right\}
 - C^{\a\b}\d_{\a\da}\d_{\b\db}\overline\eps^\da \overline\lambda^\db \nonumber \\
\overline\delta \lambda_\a &=& \overline\eps^\db C^{\c\b}
\d_{\b\db}
  \Big[i\eps_{\a\c} D + 2 \Big(F_{\a\c} + \frac{i}{2}C_{\a\c} \overline
  \lambda\overline\lambda\Big)\Big]. \label{bad1/2susytransf}
\eea
%
%
%
Take now the limit $\overline{\tau} \rightarrow \infty$. In this
limit, the action localizes to the field configuration satisfying
\bea 0 &=&  {\rm Tr} \, \epsilon^{\da\db}\left[ \overline{W}_\da
\star \overline{W}_\db
\right]_{{\bar\theta\bar\theta}} \nonumber \\
&=& {\rm Tr} \, \epsilon^{\da\db} \left[ -{1 \over
2}(\overline{\delta} \overline{\lambda}_\da) (\overline{\delta}
\overline{\lambda}_\db) + i \overline{\lambda}_\da
\overline{\delta}_\db D \right]. \nonumber \eea
Here, we have used the cyclicity of color trace and the
self-duality of the parameter $C^{\a\b}$ to simplify the last term
in the second line. Thus the partition function is localized at a
place where variations under the `broken' antichiral supersymmetry
vanish. Recall that, though ${\bar Q}$'s are broken explicitly by
the non--(anti)commutativity, {\sl linear} transformations under
the antichiral supersymmetry are well-defined. Therefore, for
infinitesimal variations, the localization is a meaningful notion.
We now see from (\ref{bad1/2susytransf}) that the localization
takes place precisely at the critical point (\ref{sdeqns}). We
will call the solutions of (\ref{sdeqns}) {\sl holomorphic
instantons}, since their amplitude is proportional to $\exp(2 \pi
i \tau)$.

For holomorphic instantons, chiral fermion zero-modes are
protected. A nontrivial solution to the $\la$ equation in
(\ref{sdeqns}) breaks the ${\cal N}={1 \over 2}$ supersymmetry
spontaneously. Therefore, these zero-modes are true Goldstino
modes, associated to the spontaneously broken ${\cal N}={1 \over
2}$ supersymmetry generated by $Q_\a$. There will be also
superconformal zero-modes, since the theory is actually invariant
under the ${\cal N}={1 \over 2}$ superconformal transformations,
part of which includes the antichiral superconformal generators
$\overline{S}^\da$. Essentially, from the viewpoint of ${\cal
N}=1$ super Yang-Mills theory, the ${\cal N}={1 \over 2}$
supersymmetry coincides with the part spontaneously broken by the
instantons.

\hfill\break

Summarizing the above considerations, antiholomorphic instantons
are solutions of the anti-self-duality equations
\bea F^{(+)}_{mn} + {i \over 2} C_{mn} \, \overline{\lambda
\lambda} = 0, \quad i \sigma^m \nabla_m \overline{\la} = 0, \quad
\la = 0, \quad {\rm Tr} {1 \over 8 \pi^2}\int_{\mathbb{R}^4}  F
\wedge F = \mathbb{Z}_- \, , \label{asdeqnsummary}\eea
while holomorphic instantons are solutions of the self-duality
equations
\bea F^{(-)}_{mn} = 0, \qquad \,\,  i \overline{\sigma}^m \nabla_m
\la = 0, \,\, \qquad \overline{\la} = 0, \,\, \qquad {\rm Tr} {1
\over 8 \pi^2}\int_{\mathbb{R}^4} F \wedge F = \mathbb{Z}_+ \, .
\label{sdeqnsummary} \eea
%

\section{Constructing Instantons for G$=$ U($2$)}
We will begin with the gauge group $G=$U(2), as in this case the
back-reaction of the fermion quasi zero-modes is rather
trivial.\footnote{This case of $G=$U(2) was also considered in
\cite{imaanpur} and \cite{grassi}.} We will always trade the U(2)
color indices for chiral or antichiral SU(2)$\times$U(1) indices,
so we express the gauge potential as
\bea\label{SpinGroup} A^{\{ab\}}_{\a\da}  \equiv (2i{\bf T}_2
A_{\a\da})_{ab}. \nonumber \eea
Of the Lie algebra $u(2)$, the symmetric part $(ab)$ realizes the
$su(2)$ subalgebra, while the antisymmetric part $[ab]$ realizes
the $u(1)$ subalgebra.

As elaborated in the previous section, the self-duality equations
(\ref{sdeqnsummary}) are exactly the same as that of ${\cal N}=1$
super Yang-Mills theory, i.e.\ these equations are not deformed by
turning on the non--(anti)commutativity. Hence the antiholomorphic
instanton solutions are the same as those of ${\cal N}=1$ super
Yang-Mills theory. For a single antiholomorphic instanton of size
$\rho$ and center $x_0$, the gauge potential and the field
strength are
\bea A^{\{ab\}}_{\b \db} = - {2 i \over [(x-x_0)^2 + \rho^2]}
\delta^{(a}_\b x^{b)}_\db, \qquad F^{\{ab\}}_{\a\b} = {8 i \rho^2
\over [(x-x_0)^2 + \rho^2]^2} \delta^{(a}_\a \delta^{b)}_\b,
\nonumber \eea
while the supersymmetry and the superconformal zero-modes $\zeta,
\overline{\eta}$ of the chiral fermion $\lambda$ (associated with
the spontaneously broken ${\cal N}={1 \over 2}$ supersymmetry) enter as
\bea \lambda_{\a} = F_{\a\b} \xi^\b, \qquad \mbox{where} \qquad
\xi^\a = \zeta^\a + x^\a_\ald \overline{\eta}^\ald \, . \nonumber
\eea
Since the anti-instanton is unaffected by the
non(anti)commutativity and does not entail any new features, we
shall not discuss it further.

The anti-self-duality equations (\ref{asdeqnsummary}) show that
the gauge field strength is modified by quasi zero-modes of
the fermion $\overline{\lambda}$. The coupled first-order
equations (\ref{sdeqnsummary}) are solvable by formally treating
the deformation parameter $C^{\a\b}$ as a perturbation and
iterating fermion back-reactions. Because of the Grassmann nature of
the fermion zero-modes, the iterative procedure will terminate,
and we will be able to construct the exact
instanton solution.

So, begin with the solution at zeroth order in $C^{\a\b}$. This is
the standard instanton solution, solving the anti-self-duality
equation, and is given by
\bea\label{ShfGauge}
A^{(0)\{ab\}}_{\beta\bed}=-\frac{2i}{[(x-x_0)^2+\rho^2]}\delta^{(a}_{\bed}
x^{b)}_\beta, \qquad
F^{(0)\{ab\}}_{\ald\bed}=\frac{8i\rho^2}{[(x-x_0)^2+\rho^2]^2}
\delta^{(a}_{\ald}\delta^{b)}_\bed. \eea
The zeroth-order
solution for the quasi zero-modes of $\overline{\la}$
(transforming as an adjoint under the SU(2) subgroup) is also
standard:
\bea\label{ShfFerm} {\bar\la^{(0)}}_\ald=
F^{(0)}_{\ald\bed}{\bar\xi}^\bed, \qquad \mbox{where} \qquad
\overline{\xi}^\bed \equiv {\overline \zeta}^\bed+
{x}_\alpha^\bed\eta^\alpha \, . \eea

In computing first-order corrections to the gauge potential, it is
useful to keep track of the color indices. For the gauge group
G$=$U(2), the bilinear $({\bar\la}{\bar\la})^{\{ab\}}$ is
antisymmetric in the color indices $a,b$ for an arbitrary spinor
${\bar\la}$, so the ${\cal O}(C)$ perturbation acts only on the
diagonal U(1) subgroup, not on the SU(2) subgroup. In particular,
one can express the perturbation as
\bea
\frac{i}{2}C_{mn}({\overline\la}^{(0)}{\overline\la}^{(0)})^{\{ab\}}=
-\eps^{ab}\frac{i}{4}C_{mn}\eps_{cd}({\overline\la}^{(0)}
{\overline\la}^{(0)})^{\{cd\}} \, . \label{u2simple} \eea
This observation is elementary but simplifies the back-reaction
computation considerably, and renders {\sl the $SU(2)$ part of the
instanton solution unaffected by the non(anti)commutativity}. On
the other hand, as we will see in the next section, this
simplification no longer works for gauge groups G$=$U($N \ge 3$).

The anti-self-duality equation of the diagonal U(1) part now
reads:
\bea (F+^*F)_{mn}=-{i\over 2} C_{mn}\,
({\overline\la\la})^{\{cd\}}\eps_{cd} \, .\nonumber \eea
To solve this equation, we first take the exterior derivative of the equation
and obtain (after using the Bianchi identity):
\bea \rmd^*F (x) =-{i\over 2}C\wedge \,
\rmd(\overline{\la\la})^{\{cd\}}(x) \eps_{cd}  \, . \nonumber \eea
This equation reduces in the Lorentz gauge to:
\bea \Box \, {}^* \! A (x) = \left[ {i \over 2} C \wedge \rmd
(\overline{\la\la})^{\{cd\}}(x)\eps_{cd} \right] \, . \nonumber
\eea
Introduce a prepotential $\Phi(x)$ for the gauge potential,
and denote the fermion bilinear as a source $J$:
\bea A (x) = {}^*[ C \wedge \rmd \Phi(x)] \qquad \mbox{and}
\qquad J(x) ={i\over 2}\eps_{cd}(\overline{\la\la})^{\{cd\}}(x).
\nonumber \eea
Notice that this prepotential ansatz for the gauge potential is
consistent with the choice of the Lorentz gauge. We have thus
reduced the first-order perturbation problem to solving a Poisson
equation:
\bea \Box \Phi (x) = J(x) \qquad {\rm where} \qquad J(x) = 3 \cdot
2^6 i\,
\frac{\rho^4}{(x^2+\rho^2)^4}{\bar\zeta}_\ald{\bar\zeta}^\ald \, .
\label{poisson} \eea
Three remarks are in order. First, it is worth emphasizing that
the above procedure applies to the construction of
multi-instantons as well. Second, concerning the field profile on
$\mathbb{R}^4$, not only is the zeroth-order solution
(\ref{ShfGauge}, \ref{ShfFerm}) SO(4) rotationally symmetric, but
the deformed solution (\ref{poisson}) is also. The SO(4) symmetry
is certainly larger than the spacetime symmetry identified in
section 2. Third, the fermion zero modes are not deformed by the
non(anti)commutativity at all. In the latter two points, the gauge
group $G=$SU(2) is exceptional. In the next two sections, for
higher-rank gauge groups, we will show that the instanton solution
is only axisymmetric, retaining symmetries belonging to SO(3)
$\subset$ SO(4) and that the fermion zero modes are deformed
further.

\section{Deformed Instantons for G$=$U($N \ge 3$): Half of the Story}
\label{SectNSC}

We next consider the gauge group G$=$U($N$) for $N \ge 3$ and find
an exact solution for the antiholomorphic instanton. We do so by
adopting the same iterative procedure, as it truncates at a finite
order in the perturbative expansion. The procedure is, however,
far more nontrivial than the G$=$U(2) case, since there are extra
fermion zero-modes. To illustrate our strategy for constructing
instantons exactly, we consider in this section a special solution
in which the superconformal quasi zero-modes are all set to zero.

Again, we start with the standard SU(2) instanton as the
zeroth-order solution\footnote{Our conventions and notations are
summarized in Appendix \ref{AppA}, and the explicit form of the
undeformed instanton solution is given in Appendix \ref{AppApr}.}
and then use perturbation theory in powers of $C^{\a\b}$ to
construct deformed solutions of (\ref{asdeqnsummary}). At
zeroth-order, the SU(2) instanton is embedded inside U($N$), so we
will decompose various U($N$) fields into U(2)$\times$U($N-2$): an
adjoint $({\bf 3}\oplus{\bf 1}, {\bf 1})$, fundamentals $({\bf 2},
\overline{\bf N \! - \! 2})$ and anti-fundamentals $(\overline{\bf
2}, {\bf N \! - \! 2})$, and singlets $({\bf 1}, {\bf N \!-\! 2}
\otimes {\bf N\!-\! 2})$ under U(2). We use the freedom of global
gauge transformation under U($N-2$) to  put the gaugino components
transforming as fundamentals to some arbitrary uni-directional
components in the U($N-2$) subspace. More precisely, we consider
the zero mode
\bea {\bar\la}^{(0)\{a\}i}_\ald=\frac{\chi^i}{(x^2+\rho^2)^{3/2}}
\delta^a_\ald\, , \nonumber \eea
and perform a U($N-2$) rotation to put $\chi^i$ in the form:
$\chi^3\ne 0$, $\chi^4=\chi^5=\dots=\chi^{N}=0$.\footnote{ From
this point onward, all equations should be interpreted as
equations in this particular frame, where $\chi_i=0$ for $i>3$.
However, for presentational purposes, we will still keep the index
$i$ for $\chi_i=0$. Notice that we do not impose any restriction
on the conjugate representation spinor ${\bar\chi}_i$.} In this
way, we have reduced the effective number of gaugino equations to
be solved in color space. Notice that the same gauge rotation does
not in general put the gauge fields to the same uni-directional
components in color space -- they are generically nonzero and need
to be solved through the anti-self-duality equations.

We expand the gauge field and fermionic zero modes in powers of
$C_{mn}$:
\bea A_m=A_m^{(0)}+A_m^{(1)}+\dots\qquad \mbox{and} \qquad
{\overline\la}_\ald={\overline\la}^{(0)}_\ald+{\overline\la}^{(1)}_\ald+\dots
\, ,  \eea
where $A_m^{(k)}$ and ${\bar\la}^{(k)}_\ald$ are of order ${\cal
O}(C^k)$, and $A_m^{(0)}$ and ${\overline\la}^{(0)}_\ald$ refer to
the undeformed single instanton solution. Normally, such an
iterative procedure would never yield an exact solution. In the
present case, what saves us is the fact that the back-reaction is
generated by a finite number of fermion zero-modes. As they are
Grassmann-valued, after some finitely many steps of the iteration,
the back-reaction terminates automatically. This is the motivation
to first consider a special solution without superconformal zero
modes, as the iteration there stops already at second order.

\subsection{First-order back-reaction}
We now solve explicitly the first-order back-reaction to the gauge
and gaugino fields. Those residing in the U($N-2$) subgroup are
not affected at all, so we only need to concentrate on the U(2)
subgroup.

First, the self-dual gauge field equation becomes:
\bea \Big(\nabla_m A^{(1)}_n-\nabla_n A^{(1)}_m
\Big)^{(+)}=-\frac{i}{2}C_{mn}{\overline\la}^{(0)}{\overline\la}^{(0)}.
\nonumber \eea
Notice that in this paper we always use $\nabla_m$ to denote a
covariant derivative with respect to the background gauge
potential $A^{(0)}_m$. A more proper notation would be
$\nabla^{(0)}_m$, but we hope that using $\nabla_m$ does not lead
to confusion. An equation of this sort can be reduced to a Laplace
equation (see the discussion in Appendix \ref{AppB}) by taking an
ansatz expressing the first-order gauge field in terms of a
matrix-valued prepotential $\Phi^{(1)}$:
\bea A^{(1)}_m (x) =C_{mn}\nabla_n \Phi^{(1)}(x), \nonumber
\eea
The resulting Laplace equation for the $(N \times N)$ matrix
prepotential $\Phi(x)$ can be easily solved:
\bea \label{SlvAone1} {\Phi^{(1)}_a}^b (x) =\delta_a^b
\Big(\phi_1(x) \,  {\overline\zeta}_\ald {\overline\zeta}^\ald+
\phi_2(x) \, {{\overline\chi}_i\chi^i \over \rho^2} \Big)
\qquad \mbox{and} \qquad {\Phi^{(1)}_i}^j=\phi_3(x) \,
{{\overline\chi}_i\chi^j \over \rho^2}, \nonumber \eea
where
\bea \phi_1 &=&-8i \left[\frac{1}{(r^2 + \rho^2)} +
\frac{\rho^2}{(r^2+\rho^2)^2} \right] \, ,  \nonumber \\
\phi_2 &=& \frac{i}{8}\frac{1}{\rho^2(r^2+\rho^2)} \, ,  \nonumber \\
\phi_3 &=& \frac{i}{4}\frac{1}{\rho^2(r^2+\rho^2)} \, .
\nonumber \eea

Next, we simplify the Weyl equation for ${\overline\la}^{(1)}$.
Substituting the value of $A_m^{(1)}$ into the equation for
${\overline\la}$, we get in the first order in $C_{mn}$:
\bea \sigma^m_{\alpha\ald}\nabla_m{\overline\la}^{(1)\ald}=
-\frac{i}{2}[A^{(1)}_m,\sigma^m_{\alpha\ald}{\overline\la}^{(0)\ald}]=
-\frac{i}{2}C_{mn}[(\nabla_n \Phi^{(1)}),
\sigma^m_{\alpha\ald}{\overline\la}^{(0)\ald}] \, .\nonumber \eea
Using the Fierz identity:
\bea\label{FierzAA}
C_{mn}\sigma^m_{\alpha\ald}=\sigma_{n\beta\ald}
C^{\beta\gamma}\eps_{\gamma\alpha}, \nonumber\eea
and the Weyl equation for $\overline{\la}^{(0)}$, one can show
that
\bea C_{mn}\sigma^m_{\alpha\ald}\nabla^n {\bar\la}^{(0)\ald}=
C^{\beta\gamma}\eps_{\gamma\alpha} \sigma_{n\beta\ald}\nabla^n
{\bar\la}^{(0)\ald}=0. \nonumber \eea
This simplifies the first-order Weyl equation for
$\overline{\la}^{(1)}$ as
\bea \sigma^m_{\alpha\ald}\nabla_m{\bar\la}^{(1)\ald}=
\frac{i}{2}{C_\alpha}^\beta\sigma^m_{\beta\ald}\nabla_m
\left[\Phi^{(1)},{\bar\la}^{(0)\ald}\right]. \nonumber \eea
We take an ansatz for $\overline{\la}^{(1)}$ in terms of a spinor
prepotential $\widehat\Psi^{(1)}$ as
\bea {\overline\la}^{(1)\bed}={\overline\sigma}^{m\bed\beta}
\nabla_m{\widehat\Psi}^{(1)}_\beta. \nonumber \eea
Using the anti-self-duality condition $F^{(0)}_{\alpha\beta}=0$, we
then get:
\bea\label{EqnXiHat} \nabla^2 \, \widehat{\Psi}^{(1)}_\alpha= -
\frac{i}{2} {C_\alpha}^\beta\sigma^m_{\beta\ald}\nabla_m
\left[\Phi^{(1)},{\overline\la}^{(0)\ald}\right]. \eea
Here, $\nabla^2\equiv \nabla_m\nabla^m$ is the covariant Laplacian
with respect to the background gauge potential $A^{(0)}_m$. We
look for the solution in a form which factorizes out the $C$
dependence from the prepotential:\footnote{This step need not
work in general, but will be justified by our explicit solution.}
\bea {\widehat\Psi}^{(1)}_\alpha={C_\alpha}^\beta
\Psi^{(1)}_\beta. \nonumber \eea
Then one can re-express (\ref{EqnXiHat})  as
\bea i{C_\alpha}^\beta\sigma^m_{\beta\ald}\nabla_m\left[
-i{\bar\sigma}^{n\ald\gamma}\nabla_n\Psi^{(1)}_\gamma-\frac{1}{2}
[\Phi^{(1)},{\overline\la}^{(0)\ald}]\right]=0. \nonumber \eea
A particular solution to this equation obeys
\bea
{\overline\sigma}^{n\ald\gamma}\nabla_n\Psi^{(1)}_\gamma=\frac{i}{2}
[\Phi^{(1)},{\overline\la}^{(0)\ald}] \, . \nonumber \eea
One can solve this equation again by using Green's functions for
the Dirac operator. We obtain the solution:
\bea\label{FermOne} {({\Psi}^{(1)}_\alpha)_a}^b &=&-{i \over 4}
\delta_a^b \frac{x_{\alpha\ald}}{\rho^2} {\bar\zeta}^\ald \left[
\frac{1}{(r^2+\rho^2)^2}\right] {\overline\chi}_k \chi^k
\nonumber \\
{({\Psi}^{(1)}_\alpha)_i}^j &=& + {i\over 2}
\frac{x_{\alpha\ald}}{\rho^2} {\bar\zeta}^\ald
\left[\frac{1}{(r^2+\rho^2)^2}\right] {\overline\chi}_i \chi^j
\nonumber\\
{({\Psi}^{(1)}_\alpha)_i}^a&=&-\frac{x_{\alpha}^{a}}{\rho^4}
{\overline\chi}_i \left[\frac{1}{(r^2+\rho^2)^{1/2}} +
\frac{\rho^4}{(r^2 + \rho^2)^{5/2}}\right]
         {\overline\zeta_\bed}{\overline\zeta^\bed},
         \nonumber \\
{({\Psi}^{(1)}_\alpha)_a}^i &=& + \frac{x_{\alpha a}}{\rho^4}
{\chi}^i \left[\frac{1}{(r^2+\rho^2)^{1/2}} + \frac{\rho^4}{(r^2 +
\rho^2)^{5/2}}\right]{\overline\zeta_\bed}{\overline\zeta^\bed}
         \, . \,\,\,
\eea
This completes the first-order computation for back-reaction of
the fermion quasi zero-modes.

\subsection{Second-order back-reaction}
Next we compute the second-order back-reaction. The second-order
perturbation for the gauge field $A_m^{(2)}$ satisfies the
equation:
\bea \left(\nabla_m A_n^{(2)}-\nabla_n A_m^{(2)}
\right)^{(+)} +\frac{i}{2}\left[A^{(1)}_m,A^{(1)}_n\right]^+
+\frac{i}{2}C_{mn}
\left({\overline\la}_\ald^{(1)}{\overline\la}^{(0)\ald}+
{\overline\la}_\ald^{(0)}{\overline\la}^{(1)\ald}\right)=0.
\label{2ndeqn} \eea
Again, we remind the readers that the superscript $(+)$ denotes
projection onto self-dual components of the antisymmetric tensor.
Begin in (\ref{2ndeqn}) with the term:
\bea \left[A^{(1)}_m,A^{(1)}_n\right]^{(+)}=
(C_{mk}C_{nl}+\frac{1}{2}\eps_{mnpq}C_{pk}C_{ql})
\nabla_{[k}\Phi^{(1)}\nabla_{l]}\Phi^{(1)} \, . \nonumber\eea
For an arbitrary antisymmetric tensor $T_{kl}$, by straightforward
computation, one finds an identity:
\bea (C_{mk}C_{nl}+\frac{1}{2}\eps_{mnpq}C_{pk}C_{ql})T_{kl}=
-\frac{1}{2}C_{kl}C_{kl}\,T^{(+)}_{mn} + C_{mn}C_{kl}\,T_{kl}.
\nonumber \eea
This allows us to simplify the commutator:
\bea \left[A^{(1)}_m,A^{(1)}_n
\right]^{(+)}=-\frac{1}{2}C_{kl}C_{kl}\left(\nabla_{[m}(\Phi^{(1)}\nabla_{n]}
\Phi^{(1)}) \right)^{(+)}
+C_{mn}C_{kl}\nabla_{[k}\Phi^{(1)}\nabla_{l]}\Phi^{(1)} \, .
\nonumber \eea
Next, express the fermion contribution in (\ref{2ndeqn}) in terms
of the prepotential $\Psi_\a^{(1)}$ in (\ref{FermOne}):
\bea {\overline\la}_\ald^{(0)}{\overline\la}^{(1)\ald}
=\frac{1}{2}C_{kl}{\overline\la}_\ald^{(0)}{\bar\sigma}^{m\ald\alpha}
{{\sigma^{kl}}_\alpha}^\beta\nabla_m \Psi_\beta^{(1)} \, .
\nonumber \eea
Using the identity
\bea
{\overline\sigma}^{m}\sigma^{kl}=\frac{1}{2}(\eta^{ml}{\overline\sigma}^k-
\eta^{mk}{\overline\sigma}^l-\eps^{mkln}{\overline\sigma}^n),
\nonumber\eea
and the self--duality of the non--(anti)commutativity tensor
$C_{kl}$, the fermion contribution in (\ref{2ndeqn}) can be
simplified as
\bea
{\overline\la}_\ald^{(0)}{\overline\la}^{(1)\ald}=C_{kl}{\overline\la}_\ald^{(0)}
{\overline\sigma}^{k\ald\alpha}\nabla_l\Psi^{(1)}_\alpha \, .
\nonumber \eea
Substituting these two expressions into (\ref{2ndeqn}) for
$A^{(2)}_m$, one finds
\bea \Big( \nabla_{[m} A_{n]}^{(2)} \Big)^{(+)}
\!\!-\frac{i}{8}C_{kl}C_{kl}\Big(
\nabla_{[m}(\Phi^{(1)}\nabla_{n]}\Phi^{(1)})\Big)^{(+)}
\!\!+\frac{i}{4}C_{mn}C_{kl}\Big(
\nabla_{[k}\Phi^{(1)}\nabla_{l]}\Phi^{(1)}+
{\overline\sigma}^{k\ald\alpha}[{\overline\la}_\ald^{(0)},\nabla_l
\Psi^{(1)}_\alpha] \Big)=0 \, . \nonumber \eea
This equation is solvable by taking again a prepotential ansatz of
the form:
\bea
A^{(2)}_m=\frac{i}{8}C_{kl}C_{kl}\Phi\nabla_m\Phi+C_{mn}\nabla_n\Phi^{(2)}
\, , \nonumber \eea
and reducing it to the Poisson equation for the $(N \times N)$
matrix-valued prepotential $\Phi^{(2)}$:
\bea \nabla^2 \, \Phi^{(2)} =iC_{kl}\left(
\nabla_{[k}\Phi^{(1)} \nabla_{l]}\Phi^{(1)} +
{\overline\sigma}^{k\ald\alpha}[{\overline\la}_\ald^{(0)},\nabla_l
\Psi^{(1)}_\alpha] \right) \, . \nonumber \eea
Again, the solution is obtained by convolving the scalar Green
function on the right-hand-side. We find that $\Phi^{(2)}$ has
nonzero components on the SU(2) subspace only:
\bea\label{SlvAone2} {(\Phi^{(2)})_a}^b
=-2iC_{mk}{({\overline\sigma}^{kn})^b}_a \left[ {x_m \, x_n \over
\rho^4} \left(\frac{1}{(r^2 + \rho^2)^2} +
\frac{\rho^2}{(r^2+\rho^2)^3} \right) \right] {\overline
\chi}_i\chi^i{\overline\zeta}_\ald{\overline\zeta}^\ald \, .\eea

To complete the iteration, one would next substitute the solution
found above into the ${\overline\la}^{(2)}_\ald$ field equations
and solve the second-order back-reaction to the fermion quasi
zero-modes. It is readily counted that the source term in the
corresponding Weyl equation contains fifth powers of the fermion
quasi zero-modes. Now that there are only four zero-modes
${\overline\zeta}_\ald$, ${\overline\chi}_i$ and $\chi^i$, the
source term vanishes identically. We thus find that second-order
back-reaction to the fermions is absent, i.e.\
${\overline\la}^{(2)}_\ald=0$. By the same reasoning, all
higher-order back-reactions ${A_m}^{(k)}$ and
${\overline\la}^{(k)}_\ald$ vanish identically for $k>2$.

In summary, in the special situation where the superconformal zero-mode
is set to zero, we have succeeded in obtaining the exact solution
for the anti-instanton as:
\bea A_m &=& A_m^{(0)}+C_{mn}\nabla_n
\Phi^{(1)}+\frac{i}{8}C_{kl}C_{kl}\Phi^{(1)} \nabla_m\Phi^{(1)}
+C_{mn}\nabla_n\Phi^{(2)}
\nonumber \\
{\overline\la}^\ald &=&
{\overline\la}^{(0)\ald}+{\overline\sigma}^{m\ald\alpha}
{C_\alpha}^\beta\nabla_m \Psi^{(1)}_\beta, \label{solwoconformal}
\eea
where the bosonic prepotentials $\Phi^{(1)}, \Phi^{(2)}$ are given
in (\ref{SlvAone1}) and (\ref{SlvAone2}), while the fermionic
prepotential $\Psi^{(1)}_\a$ is given in (\ref{FermOne}).

\section{Deformed Instantons for G$=$U($N \ge 3$): Full Story}
We now extend the result of the previous section and obtain an
exact solution for the antiholomorphic instanton in which all
quasi zero-modes of the antichiral gaugino are turned on. Compared
to the previous section, the iterative steps do not truncate at
the second-order because the superconformal zero-modes of the
fermions render the source terms in the Poisson equation far more
complicated. Nevertheless, as there are only a finite number of
fermion zero-modes, the iteration is truncated beyond some higher
order. One can thus follow the Green function method illustrated
in the previous section to  obtain the instanton solution.

We performed the computation in the way indicated and obtained the
exact anti-instanton solution in the following form:
\bea\label{ResultAns} A_m&=&A^{(0)}_m+C_{mn}\nabla_n
\Big(\Phi^{(1)}+\Phi^{(2)}+\Phi^{(3)} \Big)+
\frac{i}{16}C_{kl}C_{kl}
\Big[\Phi^{(1)}, \nabla_n\Phi^{(1)} \Big],
\nonumber\\
{\overline\la}^\ald&=&{\overline\la}^{(0)\ald}+{\bar\sigma}^{m\ald\alpha}
{C_\alpha}^\beta\nabla_m
\left(\Psi_\beta^{(1)}+\Psi_\beta^{(2)}\right)-\frac{C_{kl}C^{kl}}{32}
\Big[\Phi^{(1)}, \Big[\Phi^{(1)},{\overline\la}^{(0)\ald}\Big]
\Big] \, . \eea
Here, the superscripts denote the order of $C$-expansion they
contribute. The zeroth-order, undeformed solution $A_m^{(0)}$ is
summarized in Appendix \ref{AppApr}. Now that we have $2N$ fermion
zero-modes, reduced effectively via U($N-2$) gauge rotation to 6 zero-modes
(${\overline\zeta}_\ald$, $\eta^\a$, ${\bar\chi}_i$,
$\chi^i$), the perturbation expansions continue to the third-order
for the gauge field and to the second-order for the fermion
zero-modes. We relegate computational details to Appendix \ref{AppC}, and
collect below the final result, order by
order, using the shorthand notation $\overline{\xi}^\ald \equiv
\overline{\zeta}^\ald + x^\ald_\a \eta^\a$. The bosonic prepotentials are
\bea\label{ResultBos1} {(\Phi^{(1)})_a}^b &=&-8 i \left[
\frac{\rho^2}{(r^2+\rho^2)^2} {\overline\xi}_\ald{\overline
\xi}^\ald+
\frac{1}{r^2+\rho^2}({\overline\zeta}_\ald{\overline\zeta}^\ald+\rho^2
\eta^\alpha\eta_\alpha)- \frac{1}{r^2+\rho^2}
{{\overline\chi}_i\chi^i \over 64 \rho^2} \right] \, \delta_a^b \nonumber\\
{(\Phi^{(1)})_{ a}}^{i} &=& -\frac{2{\overline\xi}_{\dot
a}\chi^i}{(r^2+\rho^2)^{3/2}}; \quad {(\Phi^{(1)})_{i}}^{
a}=-\frac{2{\overline\chi}_i{\overline\xi}^{\dot
a}}{(r^2+\rho^2)^{3/2}}; \quad {(\Phi^{(1)})_{i}}^{j}= i {1 \over
r^2+\rho^2} \frac{{\bar\chi}_i\chi^j}{4\rho^2} \, \eea
at ${\cal O}(C)$,
\bea\label{ResultBos2}
{(\Phi^{(2)})_a}^b&=&
-2i C_{mk}{1 \over (\rho^2+r^2)^3}{({\overline\sigma}^{kn})^b}_a
{x_mx_n {\overline\chi}_i\chi^i \over \rho^2}
\left[\frac{{\overline\zeta}_\ald{\overline\zeta}^\ald}{\rho^2}(r^2+2\rho^2)-
\rho^2\eta^\alpha\eta_\alpha-\eta^\alpha
x_{\alpha\ald}{\overline\zeta}^\ald
\right]\nonumber\\
&& +2i {1 \over (\rho^2+r^2)^2}\frac{{\overline\chi}_i\chi^i}{\rho^2}
\left[ {\overline\zeta}_a
({\overline x}C)^{b\alpha}\eta_\alpha+
{\overline\zeta}^b {({\overline x}C)_a}^{\alpha}\eta_\alpha\right]
\nonumber\\
{(\Phi^{(2)})_i}^a&=&-8\frac{{\overline\chi}_i}{(r^2+\rho^2)^{5/2}}
\left[(r^2+2\rho^2){{\overline\zeta}_\ald{\overline\zeta}^\ald \over \rho^2}
({\overline x}C)^{a\alpha}\eta_\alpha+ \eta^\alpha\eta_\alpha
{({\overline x}C x)^a}_\ald{\overline\zeta}^\ald
\right]\\
{(\Phi^{(2)})_a}^i&=&-8\frac{\chi^i}{(r^2+\rho^2)^{5/2}}
\left[(r^2+2\rho^2){{\overline\zeta}_\ald{\overline\zeta}^\ald \over \rho^2}
{({\overline x}C)_a}^{\alpha}\eta_\alpha+
\eta^\alpha\eta_\alpha {({\overline
x}C x)}_{a\ald}{\overline\zeta}^\ald \right]\nonumber \eea
at ${\cal O}(C^2)$, and
\bea\label{ResultBos3} \Phi^{(3)}=i \frac{C_{kl}C_{kl}}{2}
\frac{\eta^\alpha\eta_\alpha{\bar\zeta}_\ald{\bar\zeta}^\ald{\bar
\chi}_i\chi^i}{ \rho^4(r^2+\rho^2)^3}
{\mbox{diag}}\Big(r^4+6r^2\rho^2+3\rho^4,r^4+6r^2\rho^2+3\rho^4,
2(r^4+4r^2\rho^2+\rho^4) \Big) \eea
at ${\cal O}(C^3)$, respectively. Here, $x$ denotes the matrix
$x^{\a\ald}$, $\overline x$ denotes $x^{\ald \a}$, and $C$ is a
matrix with components
\bea
{C_\alpha}^\beta=\eps_{\alpha\gamma}C^{\gamma\beta}=\frac{1}{2}C_{mn}
{(\sigma^{mn})_\alpha}^\beta \, . \nonumber \eea
The fermionic prepotentials are
\bea\label{ResultFerm1}
{(\Psi^{(1)}_\alpha)_a}^b&=&-{i \over 4}\, \delta_a^b \left[
{x_{\alpha\ald} \overline{\xi}^\ald \over \rho^2}
 \frac{1}{(r^2+\rho^2)^2}- {r^2\eta_\alpha \over \rho^4}{1 \over (r^2+\rho^2)}
 \right] \overline{\chi}_k \chi^k
\nonumber\\
{(\Psi^{(1)}_\alpha)_i}^a&=&-\frac{{x_\a}^a}{\rho^4} {\overline\chi}_i
\left[{(r^2+3\rho^2) \over (r^2 + \rho^2)^{3/2}}
{\overline\zeta}_\bed {\overline\zeta}^\bed
+ {2 \rho^4 \over (r^2 +\rho^2)^{5/2}}
{\overline\xi}_\bed{\overline\xi}^\bed \right]
-\frac{4{\bar\zeta}^a \eta_\alpha{\bar\chi}_i}{(r^2+\rho^2)^{3/2}}\\
{(\Psi^{(1)}_\alpha)_a}^i&=& {x_{\a a} \over \rho^4}
\chi^i \left[ {(r^2+3\rho^2) \over (r^2 + \rho^2)^{3/2}}
{\overline\zeta}_\bed{\overline\zeta}^\bed
+{2 \rho^4  \over (r^2 + \rho^2)^{5/2}} {\overline\xi}_\bed{\overline\xi}^\bed
\right] +
\frac{4{\bar\zeta}_a \eta_\alpha{\chi}^i}{(r^2+\rho^2)^{3/2}} \, \nonumber\\
{(\Psi^{(1)}_\alpha)_i}^j&=&\frac{i}{2}
\frac{{\bar\chi}_i \chi^j}{\rho^2(r^2+\rho^2)}\left[
x_{\alpha\ald}{\bar\xi}^\ald-\frac{r^2}{\rho^2}\eta_\alpha\right]\, .\nonumber
\eea
at ${\cal O}(C)$ order, and
\bea\label{ResultFerm2}
{(\Psi^{(2)}_\alpha)_a}^b&=&
\frac{{\overline\chi}_i\chi^i}{\rho^2(r^2+\rho^2)^3}\Big[
4 C_{mn} x_m{({\overline\sigma}^{nk})^b}_a
x_k x_{\alpha\ald}{\overline\zeta}^\ald \, \eta^\beta\eta_\beta
-2 \Big( x_{\alpha a}{\overline x}^{b\beta}{C_\beta}^\gamma\eta_\gamma
+ {\overline x}^{b\beta}C_{\beta\alpha}\eta^\gamma x_{\gamma a} \Big)
{\overline\zeta}_\ald{\overline\zeta}^\ald
\nonumber\\
&&\hskip2.3cm -
\frac{r^2+\rho^2}{\rho^2}
\Big( x_{\alpha a}{\overline x}^{b\beta}{C_\beta}^\gamma\eta_\gamma
+
{x^b}_\a x_{\sigma a} \eta^\gamma{C_\gamma}^\sigma \Big)
{\overline\zeta}_\ald{\overline\zeta}^\ald \Big]
\nonumber \\
{(\Psi^{(2)}_\alpha)_a}^i &=&- 8 i {{C_\a}^\b x_{\beta a} \chi^i
 \over (r^2+\rho^2)^{5/2}}
\eta^\gamma\eta_\gamma{\overline\zeta}_\ald{\overline\zeta}^\ald;\qquad
\qquad {(\Psi^{(2)}_\alpha)_i}^a=-8 i \frac{{C_\alpha}^\beta
{x_\beta}^a {\overline \chi}_i}{(r^2+\rho^2)^{5/2}}
\eta^\gamma\eta_\gamma{\overline\zeta}_\ald{\overline\zeta}^\ald
 \eea
at ${\cal O}(C^2)$ order.

We emphasize again the reasoning behind truncation of the
iterative process at the third order, not $N$-th order. Though
there are $2N$ fermion zero-modes, by making use of the U($N-2$)
gauge rotation, we have brought the bi-fundamental fermions $\chi,
\overline{\chi}$ to a uni-direction. Therefore, with the aid of
the gauge freedom, we have effectively reduced the independent
components of the fermion zero-modes to 6: two (would-be)
supersymmetry $\overline{\zeta}$'s, two (would-be) superconformal
$\eta$'s, and two gauge zero-modes $\chi, \overline{\chi}$. As
such, with U($N-2$) gauge orientation chosen to be
uni-directional, the iterative procedure terminates at third
order. We stress that for presentational purposes we kept an index
$i$ for the zero modes $\chi_i$, even though the results are
relevant only in the particular frame where the only
non--vanishing component is $\chi_3$.

\section{Instanton Tomography: Polarization \& Geometry of Moduli space}
Having obtained the exact solution for one antiholomorphic
instanton, we are now ready to learn aspects of semiclassical or
nonperturbative physics in ${\cal N}={1 \over 2}$ super Yang-Mills
theory. The simplest gauge-invariant quantity we would like to
study is the action functional density, which in the present case
is simply the topological charge density ${\cal F}$. Notice that
we are interested in the {\sl density} of the topological charge,
since the latter is nothing but the zeroth-moment of the former.
The zero-modes supported by the instanton span the moduli space,
which we denote as ${\cal M}$.

With particular attention to the fate of spacetime symmetries
discussed in section 2, we are primarily interested in the
5-dimensional subspace in ${\cal M}$ spanned by the instanton
center $X^{\a\da}$ and size $\rho$. The instanton density ${\cal
F}$ then depends not only on the coordinates $x^{\a\da}$ of
$\mathbb{R}^4$ but also on the coordinates of ${\cal M}$.
Therefore, one needs to examine moments of the instanton density
${\cal F}$ separately on $\mathbb{R}^4$ and on ${\cal M}$,
respectively. This is precisely what we will do in this section.
First, we will examine the profile of ${\cal F}$ on $\mathbb{R}^4$
for a fixed position on ${\cal M}$. We will then find that the
instanton charge density ${\cal F}$ contains a dipole--moment
component (in addition to the O(4)--symmetric monopole--moment
component). The dipole--moment component refers to axially
symmetric polarization of the instanton and is invariant only
under O(3) $\subset$ O(4). Second, we will examine the profile of
${\cal F}$ on ${\cal M}$ (after integrating it over
$\mathbb{R}^4$). We will compute Hitchin's information metric and
study the deformation of the geometry of ${\cal M}$. Remarkably,
we will discover that, though the metric is deformed, the volume
measure is independent of the non(anti)commutative deformation.

\subsection{Instanton density}

The topological charge density (which is the same as the action
functional density for instantons) is defined by:
\bea {\cal F}[x; Z^A] = \mbox{Tr}_{\rm U(N)} \Big(F \wedge F
\Big)_{\rm instanton} \, . \nonumber \eea
The field configurations of the instanton are functions both of
the coordinates on $\mathbb{R}^4$ and of the bosonic and fermionic
quasi zero-modes. The quasi zero-modes span the moduli space ${\cal M}$,
so we will denote coordinates on ${\cal M}$ as $Z^A = (X^{\a\da},
\rho, \eta_\a, \overline{\zeta}^\da, \chi^i, \overline{\chi}_i )$.
Therefore, the instanton density ${\cal F}$ could depend not only
on coordinates $x^{\a\da}$ of $\mathbb{R}^4$ but also on
coordinates $Z^A$ on ${\cal M}$.

Substituting the exact one-instanton solution constructed in the
previous section, after a straightforward algebra, we obtain the
action functional density as
\bea\label{InstDensity1} {\cal
F}&=&-\frac{96\rho^4}{(r^2+\rho^2)^4} \left[ 1 - {C_{kl}C_{kl}
\over \rho^2} {(r^4-6r^2\rho^2+3\rho^4) \over (r^2 + \rho^2)^2} \,
{\overline\chi}_i\chi^i{\overline\zeta}_\ald{\overline\zeta}^\ald
\nonumber \right. \nonumber \\
&& \hskip2.5cm +  \left. 2 {C_{kl}C_{kl} \over \rho^2}
{(2r^2\rho^2-3\rho^4) \over (r^2 + \rho^2)^2}\, \left( \eta^\alpha
x_{\alpha\ald}{\overline\zeta}^\ald \, {\overline\chi}_i\chi^i
+16\rho^2r^2\, {\overline\zeta}_\ald{\overline\zeta}^\ald
 \eta^\alpha\eta_\alpha
\right) \right] \, . \eea
We have shown that SO(4) Lorentz symmetry is broken explicitly on
non(anti)commutative superspace. Still, as is evident from the
spinor index contractions, the instanton density is invariant
under SO(4) rotations, {\sl provided}, in addition to $x^{\a\da}$,
all fermionic zero-modes are rotated simultaneously. Notice that,
under this SO(4) symmetry transformation, $C_{kl}$ transforms
nontrivially but $C_{kl} C^{kl}$ is invariant. We will refer to this
invariance as SO(4) (pseudo)symmetry and make further use of it
in the following subsections.

One learns from the result (\ref{InstDensity1}) that, with
non(anti)commutativity turned on, the instanton density is
deformed from the standard one by ${\cal O}(C^2)$ contributions.
Notice that, though the antiholomorphic instanton solution itself
is modified up to cubic order in the non--(anti)commutativity
parameter $C^{\a\b}$, the instanton density terminates at
quadratic order. Notice also that there is no ${\cal O}(C)$
deformation in the instanton density. These features are not due
to any delicate cancellations, but originate from SU$(2)_L$
symmetry and the Grassmann-odd nature of the fermion zero-modes.

As it stands, the result (\ref{InstDensity1}) is quite
complicated, primarily because of the last two terms involving
various combinations of fermion zero-modes. To expose further
puzzles, recall that the topological charge of the antiholomorphic
instanton, which is defined by the integral of the action
functional density ${\cal F}$ over the Euclidean space
$\mathbb{R}^4$, equals
\bea {\cal Q}_{\rm instanton} \equiv  \int_{\mathbb{R}^4} \rmd^4 x
\, {{\cal F} \over 8 \pi^2} = - 1. \nonumber \eea
It takes an integer value, though in general the integral
depends on the fermion zero-modes.  On the other hand, the
topological charge ought to be integer-valued, and hence {\sl
independent} of the fermion zero-modes whatsoever. It is also
independent of the instanton size $\rho$, but this is a well-known
result for the ordinary instanton, again from a topological argument.
What is {\sl a priori} not so obvious in the present case is that
the result is also independent of the fermion
zero-modes.\footnote{Evidently, dependence of the result on the
fermion zero-mode would lead to Grassmann-valued c-number
contribution to the topological charge. This is unphysical.} The
way this independence on fermion zero-modes comes about is highly
nontrivial: integrals over $x$ of the second and the third terms
vanish individually. We are thus led to examine tomographically
the instanton density and understand how precisely the fermionic
zero-mode dependence is distributed.

It would also be illuminating to recast the instanton density
${\cal F}$ in the context of Maldacena's gauge-gravity
correspondence \cite{maldacena}. As is well-known in the context
of 5-dimensional anti-de Sitter spacetime as holographic dual of
${\cal N}=4$ super Yang-Mills theory, the instanton density ${\cal
F}$ defines the bulk-to-boundary propagator (as introduced in
\cite{GKPW}) of a massless bulk scalar field that couples to the
topological charge density of the super Yang-Mills theory residing
at the boundary \cite{balasubramanian}. This can be understood
from the elementary observation that
\bea \Delta^{(5)}_{\rm AdS} {1 \over 8 \pi^2} {\cal F}(Z;
x)\Big|_{C = 0} &=& 0 \hskip3cm \mbox{for} \qquad \qquad \rho \ne
0
\nonumber \\
\lim_{\rho \rightarrow 0} \, {1 \over 8 \pi^2} {\cal F}(Z; x)
\Big|_{C=0} &=& -\delta^{(4)}(x) \qquad \mbox{obeying} \qquad
\int_{\mathbb{R}^4} {{\cal F} \over 8 \pi^2} = -1.
\label{adsrelation} \eea
In this context, the coordinates $(X^{\a\da}, \rho)$ are
interpreted as the bulk location, while the coordinate $x$ refers
to the boundary location. Once the non(anti)commutativity is
turned on, neither of the two relations would hold. Therefore, one
expects that both the geometry of the 5-dimensional gravity
background and the instanton density would be modified. In the
following subsections, we will explore aspects of these
modifications in detail.

\subsection{Instanton polarization by non(anti)commutativity}
With a fair amount of guesswork based on underlying symmetries, we
were able to show that the action functional density can be
packaged into the following form:
\bea {\cal F}&=&
96(\rho^4-C^2\rho^{-2}{\bar\chi}_i\chi^i{\bar\zeta}_\ald{\bar\zeta}^\ald
+64C^2\rho^2{\bar\zeta}_\ald{\bar\zeta}^\ald\eta^\alpha\eta_\alpha)
\label{InstDensity2} \\
&\times& \left[\Big(x+\sqrt{10}C w-C^2\frac{{\bar\chi}_i\chi^i \,
w}{2\rho^4}\Big)^2+ \rho^2 +2\sqrt{10}C
\Big({\bar\zeta}_\ald{\bar\zeta}^\ald
-\frac{{\bar\chi}_i\chi^i}{16} \Big)+
2C^2{\bar\zeta}_\ald{\bar\zeta}^\ald
\Big(18\eta^\alpha\eta_\alpha-
\frac{{\bar\chi}_i\chi^i}{\rho^4}\Big)
\right]^{-2} \nonumber\\
&\times& \left[\Big(x-\sqrt{10}C w-C^2\frac{{\bar\chi}_i\chi^i\,
w}{2\rho^4} \Big)^2+ \rho^2- 2\sqrt{10}C
\Big({\bar\zeta}_\ald{\bar\zeta}^\ald
-\frac{{\bar\chi}_i\chi^i}{16}\Big)+
2C^2{\bar\zeta}_\ald{\bar\zeta}^\ald \Big(18\eta^\alpha\eta_\alpha
- \frac{{\bar\chi}_i\chi^i}{\rho^4}\Big) \right]^{-2}. \nonumber
\eea
Here, we have introduced the following shorthand notation:
\bea w^{\a\ald} \equiv \eta^\alpha{\overline\zeta}^\ald,\qquad C^2
\equiv {1 \over 4} \det C^{\a\b},\qquad  C\equiv\sqrt{C^2}.
\nonumber \eea
Notice that, in (\ref{InstDensity2}), the two square-brackets are
exchanged by the inversion $\Pi$ in $\mathbb{R}^4$ (which is the
Euclidean version of the combined operation of parity $P$ and
time-reversal $T$); since $\overline{\lambda}$ is an antichiral
fermion, $\Pi$ essentially rotates all the fermion zero-modes by
$e^{ i \pi/2} = +i$. Therefore, the new expression
(\ref{InstDensity2}) of the instanton density exhibits the
$\mathbb{Z}_2$ antipodal reflection symmetry manifestly! This $
\mathbb{Z}_2$ reflection is nothing but a subgroup of the (pseudo)
SO(4) symmetry discussed below (\ref{InstDensity1}).

In passing, we would like to emphasize that, though each
square-bracket in (\ref{InstDensity2}) seems to contain a
nonanalytic expression of $C^{\a\b}$, the instanton density ${\cal
F}$ is actually analytic -- (\ref{InstDensity2}) is merely
rewriting (\ref{InstDensity1}), whose expression is manifestly
analytic in $C^{\a\b}$.

The alternative expression (\ref{InstDensity2}) for the instanton
density now offers an intuitive understanding of the effect of
non(anti)commutativity. From the two square brackets in
(\ref{InstDensity2}), one readily finds a variety of deformations.
A class of deformation of most interest to our discussion is the
one arising from the last term in the round bracket, proportional
to $C^2 \overline{\chi}_i \chi^i w$. We will now argue that this
term corresponds to polarizing ${\cal F}$ so that the
dipole-moment is induced. \footnote{We also note that all
deformation terms in (\ref{InstDensity2}) other than $\Delta_p$
contribute to deformation of the monopole-moment component in the
instanton density ${\cal F}$.}

Begin by noting that the term proportional to $C^2 \chi^i
\overline{\chi}_i w^{\a\da}x_{\a\da}$ flips sign under the
aforementioned antipodal $\mathbb{Z}_2$ reflection. Bearing in
mind that $\mathbb{Z}_2$ is nothing but a subgroup of SO(4)
(pseudo)symmetry, one finds that the instanton density ${\cal F}$
is polarized along the direction set by
\bea \Delta_{\rm p}^{\a\da} = {C^2 \overline{\chi}_i \chi^i \over
\rho^4} w^{\a\da}. \label{dipole} \eea
As the dipole moment $\Delta_p$ is proportional to $C^2$, we
discover that the first moment of ${\cal F}$ is induced as a
consequence of turning on non(anti)commutativity --- an important
indication that the Grassmann-even space $\mathbb{R}^4$ and the
Grassmann-odd space spanned by $(\th^1, \th^2)$ are not merely in
a direct product, but rather intertwined. This is as expected. We
have shown in section 2 that the conformal extension of the
non(anti)commutative superspace is nothing but ${\cal N}={1 \over
2}$ conformal superspace. We will make this more concrete in the
next two subsections.

Notice that the dipole moment $\Delta_p$ is also set by the
product of supersymmetry and superconformal zero-modes {\sl and}
to the off-diagonal zero-modes $\chi^i, \overline{\chi}_i$. In
particular, dependence on the latter is quite interesting since
$\chi^i, \overline{\chi}_i$ zero-modes are the ones present only
for higher-rank gauge groups $G=$U($N\ge 3)$. It also follows that
the instanton in U(2) gauge theory does not support enough
fermionic zero-modes to exhibit the full intricacy of physics on
the non(anti)commutative superspace. In fact, for $G=$U(2), the
instanton density ${\cal F}$ in (\ref{InstDensity1}) is
considerably simplified and one observes
%
%
the maximum at $x^{\a\da} = 0$. It is also SO(4) rotationally
symmetric and thus carries no first moment.

After all, polarization of the antiholomorphic instanton is fully
consistent with symmetries of the non(anti)commutative superspace.
As explained in section 2, the non(anti)commutativity breaks the
underlying SO(4) Lorentz symmetry to the antichiral SU$(2)_R$
symmetry, acting on antichiral, dotted indices. This implies that
the instanton configuration would no longer be a spherically
symmetric configuration on $\mathbb{R}^4$, but rather a
configuration invariant only under SU$(2)_R$. Indeed, we have just
observed that the instanton is polarized such that its topological
charge density is axi-symmetric, where the polarization direction
is set by the product of supersymmetry and superconformal
zero-modes.

We should however emphasize that the induced polarization is set
entirely by the fermion zero-modes and hence Grassmann-valued.
Modulo this point, what underlies the polarization is precisely
the same physics as the UV/IR mixing phenomenon discovered in
\cite{MinwSeib} and by now well understood in terms of open Wilson
lines \cite{owls,leshouchesreview} in the noncommutative
spacetime. We recall that the UV/IR mixing phenomenon was also
shown to take place in non(anti)commutative superspace
\cite{brittofengrey2}. With such a caveat, we plot the induced
dipole-moment component of the instanton density in Fig. 1, and
contrast it with the monopole-moment component.

\begin{figure}
\begin{tabular}{ccccc}
\epsfysize=1.9in \epsffile{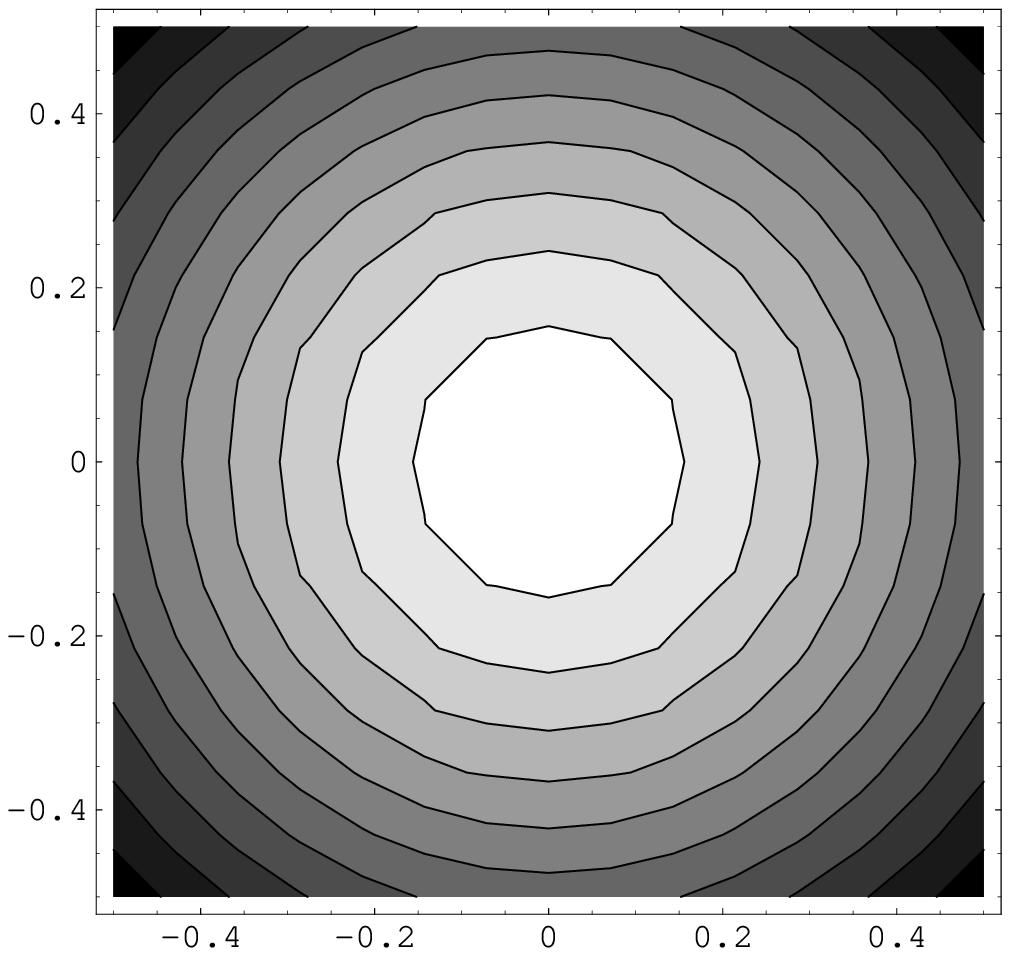}&
\begin{picture}(10.00,10.00)
\end{picture}&
\epsfysize=1.9in \epsffile{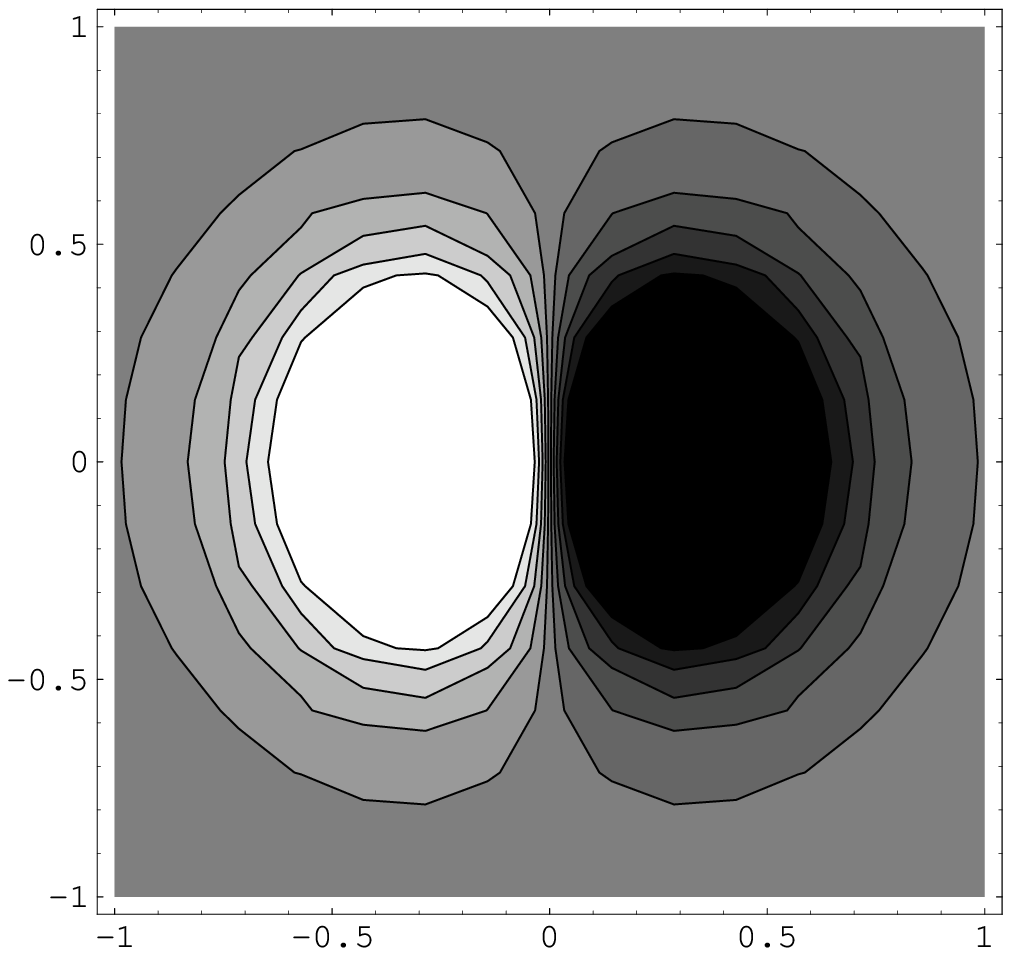}&
\begin{picture}(10.00,10.00)
\end{picture}&
\epsfysize=1.9in \epsffile{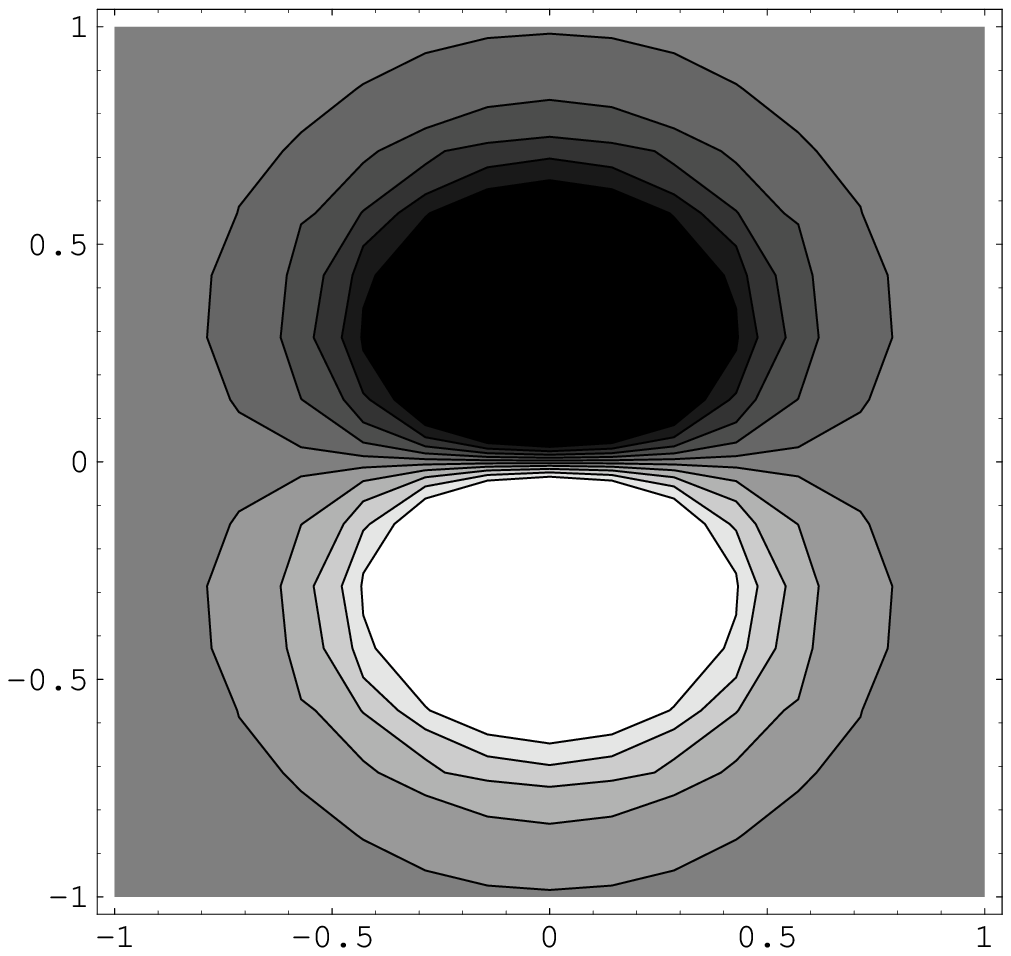}\\
(a)&&(b)&&(c)
\end{tabular}
\caption{
Contour plots of the instanton density
(\ref{InstDensity1}). We consider the simplest case where
${\bar\zeta}^{\dot 2}=\eta^2=0$ and present the section in
$x_3$--$x_4$ plane: (a) section of the $C$--independent,
monopole-moment component of (\ref{InstDensity1}); (b)
dipole-moment component for real part of coefficient of
$C^2\eta^1{\bar\zeta}^1{\bar\chi}_i\chi^i$ ; (c) dipole-moment
component for imaginary part of coefficient of
$C^2\eta^1{\bar\zeta}^1{\bar\chi}_i\chi^i$.
}
\end{figure}

\subsection{Geometry of moduli space: information metric}

We will now dwell on the other side of the instanton density
(\ref{InstDensity1}, \ref{InstDensity2}), viz. variation of the
density over one-instanton moduli space ${\cal M}$. Traditionally,
the instanton moduli space is defined in terms of the so-called
$L^2$-metric --- the induced metric on the space of zero-modes
obtained by choosing a conformal structure on
$\mathbb{R}^4$.\footnote{For a compact hyperk\"ahler manifold, it
is well-known that the $L^2$-metric on the moduli space is again
hyperk\"ahler. This assertion extends also to noncompact
hyperk\"ahler manifolds such as $\mathbb{R}^4$.} Practically, the
$L^2$-metric is not so convenient, since it comes with various
technical complications. For example, the formalism is not
manifestly gauge invariant, the moduli space is typically
afflicted with small instanton singularities at a finite distance
in the moduli space,\footnote{This is particularly a drawback for
making contact with Maldacena's gauge-gravity correspondence
\cite{maldacena}, since differential geometry is ill-defined at
the singularity.} and the metric does not exhibit manifest
conformal invariance though the (anti)-self-dual equation does. To
remedy these shortcomings, Hitchin proposed an alternative
definition of the moduli space metric based on the so-called
information metric \cite{Hitchin,InformMetr,narain}. The idea is
that one views the instanton action density as a family of
probability distributions on $\mathbb{R}^4$, parametrized by the
Grassmann-even and Grassmann-odd zero-modes of instantons.
Implicit to the idea is an assumption that the moduli space is a
submanifold of the infinite-dimensional affine space of all {\sl
smooth} volume forms with unit volume. Since we are more
interested in incorporating all the spacetime symmetries inherent
to the theory and studying differential geometry on the moduli
space, we prefer studying the information metric over the
$L^2$-metric.

As mentioned, Hitchin's information metric is defined entirely in
terms of the instanton density ${\cal F}$, and is given by
\bea\label{DefInfMetr} {\cal G}_{ab}\, \rmd Z^a \rmd Z^b\equiv
\rmd Z^a \rmd Z^b\int_{\mathbb{R}^4} \rmd^4 x \, {\d_a {\cal F}
\d_b {\cal F} \over {\cal F}} \, . \eea
The information metric has many virtues compared to the
$L^2$-metric. First, as the instanton density ${\cal F}$ is
gauge-invariant, the metric defines the moduli space geometry with
manifest gauge-invariance. Second, the metric is geodesically
complete. Third, elementary scaling analysis proves that the
metric exhibits manifest conformal invariance.

In this section, we shall compute Hitchin's information metric
explicitly for a single antiholomorphic instanton, and learn the
geometry of the moduli space ${\cal M}$. Recall that the moduli
space ${\cal M}$ is five-dimensional (apart from the trivial SU(2)
$\subset$ U(2) gauge orientations), spanning the instanton's size
and center. To introduce the instanton center position, we first
shift the coordinates $x^{\a\ald}$ on $\mathbb{R}^4$ in
(\ref{InstDensity1}) by $(x- X)^{\a\ald}$, where $X^{\a\ald}$ now
refers to the center of the anti-instanton, and compute the
integrals (\ref{DefInfMetr}). After some algebra, we find that
\bea\label{InformMetric} {\cal G}_{AB} \, \rmd Z^A \rmd
Z^B&=&\frac{2}{5}\left[ 64\frac{\rmd\rho^2}{\rho^2}\left(1-
\frac{15}{7\rho^6}C^2 S_1 -\frac{320}{7\rho^2}C^2 S_2\right)
+64\frac{\rmd X^2}{\rho^2} \left(1 -\frac{5}{7\rho^6}C^2 S_1
-\frac{88}{7\rho^2}C^2S_2\right)\right.\nonumber\\
&&\ \left. + {32C^2\over 7 \rho^7} T^m \rmd X_m \rmd\rho
-\frac{16C^2}{\rho^6}\rmd X_m \rmd T^m +\frac{32C^2}{\rho^7}
\rmd\rho \rmd S_1+ \frac{1024C^2}{\rho^3}\rmd \rho \rmd S_2\right]
\, . \eea
Here, we introduced the shorthand notation $S_1$, $S_2$ and $T_m$
for three independent products involving the fermion zero-modes:
\bea S_1\equiv
{\overline\zeta}_\ald{\overline\zeta}^\ald{\overline\chi}_i\chi^i,\qquad
S_2\equiv
{\overline\zeta}_\ald{\overline\zeta}^\ald\eta^\alpha\eta_\alpha,\qquad
T_m\equiv {\overline\chi}_i\chi^i\eta^\alpha
\sigma^m_{\alpha\ald}{\overline\zeta}^\ald \, . \eea
As they are quartic in Grassmann-odd variables, the product of
any two such objects vanishes identically.

Much as the instanton density itself, the information metric
(\ref{InformMetric}) is quite complicated because of $C$-dependent
fermionic zero-mode effects. This is again the manifestation that
the moduli space ${\cal M}$ is not the standard superconformal
superspace, but rather its non(anti)commutative counterpart with
${\cal N}={1 \over 2}$ supersymmetry explained in section 2.
Nontrivial mixing between Grassmann-even and Grassmann-odd
coordinates on ${\cal M}$ as observed in (\ref{InformMetric})
originates from the non(anti)commutativity. All these mixings,
however, are removable by a suitable change of variables.
Introduce the following shift to the Grassmann-even coordinates
$X_m$ and $\rho^2$:
\bea {\widetilde X}^m=X^m-\frac{C^2}{8\rho^4}T^m\qquad \mbox{and}
\qquad {\widetilde \rho}^2=\rho^2\left[ 1 +\frac{C^2}{2}
\frac{S_1}{\rho^6}+16C^2 \frac{S_2}{\rho^2} \right] \, .\nonumber
\eea
Notice that we still maintain the translational invariance on
conformal slice of $\mathbb{R}^4$ in ${\widetilde X}^m$. In terms
of the shifted variables,  the information metric becomes:
\bea {\cal G}_{AB}\rmd Z^A \rmd Z^B &=&\frac{128}{5}\left[
\frac{\rmd{\widetilde\rho}^2}{{\widetilde\rho}^2}\left(1+
\frac{6}{7{\tilde\rho}^6}C^2 S_1 -\frac{96}{7{\tilde\rho}^2}C^2
S_2\right) \right. \nonumber \\
&& \hskip0.4cm \left. + \frac{\rmd{\widetilde
X}^2}{{\widetilde\rho}^2} \left(1
-\frac{3}{14{\widetilde\rho}^6}C^2 S_1
+\frac{24}{7{\widetilde\rho}^2}C^2S_2\right)
-C^2\frac{13}{14{\widetilde\rho}^7}T^m \rmd{\widetilde\rho} \,
\rmd X_m  \right]. \label{newform} \eea
One readily sees that, in the limit that the Grassmann-odd
coordinates vanish, the information metric is precisely the metric
of the hyperbolic space $\mathbb{H}_5$, describing the
five-dimensional Euclidean anti-de Sitter (AdS) space. Therefore,
for nonzero $C^{\a\b}$, one would still interpret the instanton
density ${\cal F}$ as the bulk-to-boundary propagator obeying
\bea \widetilde\Delta^{(5)}_{\rm H} \, {1 \over 8 \pi^2} {\cal
F}(Z; x) &=& 0 \hskip3cm \mbox{for} \qquad \qquad Z^A \ne 0
\nonumber \\
\lim_{Z^A \rightarrow 0} \,\, {1 \over 8 \pi^2} {\cal F}(Z; x) &=&
- \delta^{(4)}(x) \hskip1.5cm \mbox{obeying} \qquad
\int_{\mathbb{R}^4} {{\cal F} \over 8 \pi^2} = - 1,
\label{newrelation} \eea
where $\widetilde\Delta^{(5)}_{\rm H}$ refers to the scalar
Laplacian operator defined by Hitchin's information metric
(\ref{InformMetric}), and the limit $Z^A \rightarrow 0$ is
interpreted as taking all Grassmann-even and Grassmann-odd
coordinates approaching zero while holding $C^{\a\b}$ finite.

A quantity of interest is the volume measure of Hitchin's
information metric, as it defines the sum over instanton effects
in ${\cal N}={1 \over 2}$ super Yang-Mills theory. From
(\ref{newform}), one readily finds that
\bea \rmd \mbox{Vol} = \sqrt{\mbox{det}{\cal G} } \, \rmd
\widetilde \rho \, \rmd^4 X \qquad \mbox{\rm where} \qquad
\mbox{det} {\cal G} = \left( {128 \over  5} \right)^5
\frac{1}{{\tilde\rho}^{10}} \, . \nonumber \eea
Remarkably, we find that the volume form is exactly the same as in
ordinary Yang-Mills theories! It should be noted however that the
natural coordinate is not $\rho$ but $\widetilde{\rho}$. This may
indicate that $\widetilde{\rho}$ should be interpreted as the size
modulus of the deformed instanton.

\subsection{Symmetry Considerations}
Having examined the instanton density and the information metric,
one might be able to infer the results from considerations of the
spacetime conformal symmetry identified in section 2. Indeed, the
functional form of various terms in the information metric
(\ref{InformMetric}) (but not the coefficients) is determinable by
those symmetries. Let us briefly discuss these symmetries and
their consequences for the metric (\ref{InformMetric}).
\hfill\break
$\bullet$ Translational invariance on the conformal slice of
$\mathbb{R}^4$ guarantees that the metric components ${\cal
G}_{ab}$ do not depend upon the translational bosonic zero-modes
$X_m$'s. \hfill\break
$\bullet$ Global rotation in the $U(N-2)$ part of the gauge group
(i.e. unitary rotations acting by $\chi^i\rightarrow
{U^i}_j\chi^j$, ${\bar\chi}_i\rightarrow {U_i}^j {\bar\chi}_j$)
restricts possible contributions of the bi--fundamental fermionic
zero--modes to ${\bar\chi}_i\chi^i$ or $\rmd
({\bar\chi}_i\chi^i)$. \hfill\break
$\bullet$ Chiral SU$(2)_L$ and antichiral SU$(2)_R$ symmetries
(acting on dotted and undotted indices) restrict possible
contractions of supersymmetric and superconformal fermionic
zero--modes to ${\overline\zeta}_\ald{\overline\zeta}^\ald$,
$\eta^\alpha\eta_\alpha$, $\eta^\alpha
\sigma^m_{\alpha\ald}{\overline\zeta}^\ald$ and
\bea\label{DropSymmetry} \eta_\alpha
C^{\alpha\beta}\sigma^m_{\beta\ald}{\bar\zeta}^\ald,\qquad
\eta^\alpha C_{\alpha\beta}C^{\beta\gamma}
\sigma^m_{\gamma\ald}{\overline\zeta}^\ald \eea
Here, we also made use of the fermionic statistics and of the fact
that we can have at most two $C_{\alpha\beta}$ if only one $\eta$
and one $\zeta$ are present. Notice that $\eta^\alpha
\sigma^m_{\alpha\ald}{\bar\zeta}^\ald$ can appear only in the
combination $\rmd X_m\eta^\alpha
\sigma^m_{\alpha\ald}{\bar\zeta}^\ald$ in order to be consistent
with Lorentz (pseudo) symmetry and translation symmetry.
\hfill\break
$\bullet$ The first term in (\ref{DropSymmetry}) is removable in
the information metric. \footnote{The second term can be written
in terms of $\eta^\alpha
\sigma^m_{\alpha\ald}{\overline\zeta}^\ald$ due to relation
$C_{\alpha\beta}C^{\beta\gamma}=
\frac{1}{2}C_{\sigma\beta}C^{\beta\sigma}\delta_\alpha^\gamma$.}
To show this, we notice that there are only two ways of
incorporating the first term in (\ref{DropSymmetry}) consistently
with Lorentz (pseudo)symmetry:
\bea\label{TempProof} \rmd X_m \eta_\alpha
C^{\alpha\beta}\sigma^m_{\beta\ald}{\overline\zeta}^\ald, \qquad
C_{mn} \rmd X_n \eta_\alpha
C^{\alpha\beta}\sigma^m_{\beta\ald}{\overline\zeta}^\ald \, . \eea
The first term is linear in $C^{\a\b}$ and thus cannot appear in
the instanton density ${\cal F}$ or in the information metric.
Recall that the part of the solution that is linear in $C$ is a
singlet under SU$(2)$. Using the Fierz identity (\ref{FierzAA}),
\bea C_{mn}\sigma^m_{\beta\ald}= \sigma_{n\rho\ald}
C^{\rho\gamma}\eps_{\gamma\beta}, \nonumber \eea
we observe that the second term in (\ref{TempProof}) becomes
\bea C_{mn} \rmd X_n \eta_\alpha
C^{\alpha\beta}\sigma^m_{\beta\ald}{\overline\zeta}^\ald= \rmd
X_n\eta_\alpha C^{\alpha\beta}\sigma_{n\rho\ald}
C^{\rho\gamma}\eps_{\gamma\beta}{\overline\zeta}^\ald=
-\frac{1}{2}C^{\rho\beta}C_{\rho\beta}dX_n
\eta^\alpha\sigma^n_{\alpha\ald}{\overline\zeta}^\ald. \nonumber
\eea
Putting these considerations together, we find that only the
following combinations of zero-modes are permitted by the
spacetime symmetries:
\bea S_1\equiv
{\overline\zeta}_\ald{\overline\zeta}^\ald{\overline\chi}_i\chi^i,\quad
S_2\equiv
{\overline\zeta}_\ald{\overline\zeta}^\ald\eta^\alpha\eta_\alpha,\quad
S_3\equiv {\eta}^\alpha{\eta}^\alpha{\overline\chi}_i\chi^i,\quad
\rmd X_m T_m\equiv \rmd X_m{\overline\chi}_i\chi^i\eta^\alpha
\sigma^m_{\alpha\ald}{\overline\zeta}^\ald \, , \nonumber \eea
and they should be multiplied by $C^2$ in the information metric
to be consistent with U(1) (pseudo) R-symmetry. \hfill\break
$\bullet$ The fermion zero-modes transform under the ${\cal N}={1
\over 2}$ supersymmetry as follows:
\bean (\delta X)_{\alpha \D \alpha} & = & 4i
\epsilon_{\alpha} \O
\zeta_{\D \alpha}, \\
\delta \rho^2 & = & 4i \rho^2 (\epsilon \eta),\\
\delta \eta & = & 4i \eta  (\epsilon \eta),\\
\delta \chi & = & 6i \chi (\epsilon \eta),\\
\delta \O \chi & = & 6i \O \chi (\epsilon \eta),\\
\delta \overline{\zeta} & = & 0. \nonumber \eean
From these rules, one readily finds that
\bea \delta \left({S_1\over \rho^6}\right)=0; \qquad \delta
\left({S_2\over \rho^2} \right)=0; \qquad {\rm and} \qquad \delta
\left( \O \chi \chi \eta^2 \right)=0. \nonumber \eea
Both the instanton density ${\cal F}$ and the information metric
${\cal G}_{AB}$ are invariant under these transformations.
\footnote{There is a possible combination of the form
$C^2S_3=C^2{\bar\chi}_i\chi^i\eta^\alpha\eta_\alpha$ that is
consistent with all symmetries; thus it can in principle appear in
the instanton density ${\cal F}$ and in the information metric
${\cal G}_{AB}$. On the other hand, explicit computation indicates
that the coefficient of this term is zero.} \hfill\break
$\bullet$ Powers of $\rho$ in the instanton density ${\cal F}$ and
the information metric ${\cal G}_{AB}$ are determinable by
elementary dimensional analysis.

\section*{Acknowledgements}
We thank D. Berenstein, J. Maldacena and N. Seiberg for helpful
discussions. RB thanks the Aspen Center for Physics for
hospitality during this work. SJR was a Member at the Institute
for Advanced Study during this work. He thanks the School of
Natural Sciences for hospitality and for the grant-in-aid from the
Fund for Natural Sciences.


\appendix

\section{conventions and notation}
\label{AppA}
The conventions and notation we adopt are as follows. The signature of
Lorentzian spacetime $\mathbb{R}^{3,1}$ is diag.$(+---)$. Wick
rotation to Euclidean $\mathbb{R}^4$ is achieved by $x^0
\rightarrow i x^4$ and $S_{\rm Lorentzian} \rightarrow i S_{\rm
Euclidean}$. We continue adopting Lorentzian spinor notation. We
freely change between SO(4) and SU$(2)_L \times$SU$(2)_R$
indices, viz.
\bean
& & \d_{\a\da} = \sigma^m_{\a\da} \d_m; \qquad
\d_{\a\da} x^{\b\db} = -2 \delta^\b_\a \delta^\db_\da\\
& & x_{\a\da} = \sigma^m_{\a\da} x_m, \qquad x_m = -\frac{1}{2}
\bar\sigma_m^{\da\a} x_{\a\da}.  \nonumber \eean
We normalize the gauge covariant derivatives as
\bean \nabla_{\a\da} = \partial_{\a\da} + {i \over 2} [A_{\a\da},
\,\,\,]. \eean
For the U(2) gauge group, we also freely interchange color indices
between adjoint and spinor indices \cite{shifman} (suppressing
spacetime indices):
\bean A := A^a {\bf T}^a ; \qquad A^{\{ab\}}:=(2i {\bf T}_2
A)^{ab}; \qquad {\rm tr}_{u(2)} {\bf T}^a {\bf T}^b = {1 \over 2}
\delta^{ab}. \eean
In spinor notation, the traceless SU(2) subgroup is symmetric
in the spinor indices $a,b$, while the diagonal U(1) subgroup is
proportional to $\eps^{ab}$. In explicit form,
\bean A^{\{ab\}}=\eps^{a{ c}}{A_{ c}}^b,\qquad {A_{ a}}^b=\eps_{{
a}c}A^{\{cb\}}. \eean
In differential forms, the gauge field strengths are
\bean F = {1 \over 2} F_{mn} \rmd x^m \wedge \rmd x^n ; \qquad
{}^* F = {1 \over 2} \widetilde{F}_{mn} \rmd x^m \wedge \rmd x^n,
\eean
where $\widetilde{F}_{mn} = {1 \over 2} \epsilon_{mnpq} F_{pq}$.
Thus, ${1 \over 2} (F_{mn} F_{mn} ) = {}^*(F \wedge {}^* F)$, and
${1 \over 2} (F_{mn} \widetilde{F}_{mn}) = {}^*(F \wedge F)$.

The gauge coupling constants $\tau, \overline{\tau}$ are
customarily taken as per Lorentzian theory conventions:
\bea \tau = {\theta_{\rm YM} \over 2 \pi} + i {4 \pi \over
g^2_{\rm YM}}; \qquad \quad \overline\tau = {\theta_{\rm YM} \over
2 \pi} - i {4 \pi \over g^2_{\rm YM}}.  \nonumber \eea
In Euclidean theory, we interpret $\tau$ and $\overline{\tau}$ as
two independent, complex-valued coupling constants.

\section{single undeformed antiholomorphic instanton}
\label{AppApr}

The zeroth-order (undeformed) solution for the gauge field is
\bea
A^{(0)\{ab\}}_{\beta\bed}=-\frac{2i}{x^2+\rho^2}(\delta^a_{\bed}x^b_\beta+
\delta^b_{\bed}x^a_\beta),\qquad
F^{(0)\{ab\}}_{\ald\bed}=\frac{8i\rho^2}{(x^2+\rho^2)^2}
(\delta^a_{\ald}\delta^b_\bed+\delta^b_{\ald}\delta^a_\bed). \eea

Consider a fermion $\overline{\a}$ transforming in the adjoint
representation of U(2). The zeroth-order solution for its
zero-modes is
\bea
{\bar\la^{(0)}}_\ald=F^{(0)}_{\ald\bed}{\bar\zeta}^\bed+
F^{(0)}_{\ald\bed}{x}_\alpha^\bed\eta^\alpha
\equiv F^{(0)}_{\ald\bed}{\bar\xi}^\bed.
\eea
Consider next fermions $\chi, \overline{\chi}$ transforming in
the fundamental and anti-fundamental representations of U(2),
respectively. The zeroth-order solution for their zero-modes
is
\bea {{\overline\la^{(0)}_{\ald a}}}~^{i}\equiv \eps_{{
a}c}{\overline\la^{(0)\{c\}i}}_{\ald}, \qquad
{\overline\la^{(0)\{a\}i}}_{\ald}=\frac{\chi^i}{(x^2+\rho^2)^{3/2}}
\delta^{a}_{\ald}, \qquad {{\overline\la^{(0)}}_{\ald i}}~^{
a}\equiv {{\overline\la^{(0)\{  a\}}}_{\ald i}}=
\frac{{\overline \chi}_i}{(x^2+\rho^2)^{3/2}}\delta^{
a}_{\ald} \nonumber \eea
The fermion zero-modes have the following scaling dimensions and
R-charges:
\begin{center}
\begin{tabular}{|c|r|c|} \hline
 & dim & U$(1)_R$ \\ \hline
$X_m$  &  $-1$  & $0$ \\ \hline
$\rho$  &  $-1$  & $0$ \\ \hline
$\eta_\alpha$ &$\frac{1}{2}$& $-1$\\ \hline
${\bar\zeta}^\ald$&$-\frac{1}{2}$& $-1$\\ \hline
${\chi}^i$&$-\frac{3}{2}$& $-1$\\ \hline
${\bar\chi}_i$&$-\frac{3}{2}$& $-1$\\ \hline
\end{tabular}
\end{center}

In the instanton solution for gauge group G$=$U(N),
$\chi_i$ is the fermion component transforming as a bi-fundamental
under U$(N-2)$, and $\overline\chi_i$ is its complex conjugate.
Starting with an arbitrary zero-mode, one can always perform a
constant U$(N-2)$ rotation, so that there is only one nontrivial
component $\chi_3$. In other words, one can reduce the general
discussion of G=U$(N)$ with $N\geq 3$ effectively to G=U$(3)$.
%
\section{solving differential equations}
\label{AppB}
While constructing the instanton solution in perturbation theory,
we repeatedly encounter equations of the following form:
\bea\label{AppBEqnStart} \nabla_m
A_n-\nabla_nA_m+\eps_{mnkl}\nabla_kA_l=-C_{mn}J \, , \eea
where the covariant derivative $\nabla_n$ is computed with instanton
background field $A^{(0)}$ taken in the appropriate
representation of $SU(2)$.
Applying a differential operator $\eps_{mnrs}\nabla_s$ to both sides
of the above equation, we find:\footnote{To simplify notation, we
use $F_{mn}$ instead of $F^{(0)}_{mn}$ in this appendix.}
\bean 2i F_{nr}A_n+2\nabla_r\nabla_l A_l-2\nabla^2
A_r=-\eps_{mnrs}\nabla_s C_{mn} J \, .\eean
Here, we used the definition of $F_{mn}$ and its
anti-self-duality:
\bea \nabla_{m}\nabla_{n}-\nabla_{n}\nabla_{m}=\frac{i}{2}
F_{mn},\qquad \eps_{mnrs}F_{sm}=2F_{nr}. \nonumber \eea
In the Lorentz gauge ($\nabla_l A_l=0$), we have:
\bean \label{AppBEqn} i F_{nm}A_n-\nabla^2 A_m=-C_{mn}\nabla_n J.
\eean
This equation is solvable by taking an ansatz:
\bea\label{AppBAnz} A_m=C_{mn}\nabla_n\Phi, \eea
which automatically satisfies the gauge condition $\nabla_m
A_m=\frac{i}{4}C_{mn}F_{mn}\Phi=0$ because of the
anti-self-duality of $F_{mn}$. Using the relation
$F_{nm}C_{nk}=F_{nk}C_{nm}$, the left-hand side of (\ref{AppBEqn})
can be rewritten as
\bean iF_{nm}A_n-\nabla^2 A_m
=-\frac{i}{2}[\nabla_k,F_{kl}]C_{ml}\Phi-C_{ml}\nabla_l\nabla^2\Phi.
\eean
We thus demonstrated that there exists a natural ansatz
(\ref{AppBAnz}) for the gauge potential $A_m$, allowing us to reduce
(\ref{AppBEqnStart}) to a {\sl single} differential equation of
Poisson type:
\bea \nabla^2\Phi= J. \nonumber \eea
Here, $\nabla$ is a covariant derivative in an appropriate
representation of the gauge group G.

\section{details of the computation}
\label{AppC} In this appendix we present some intermediate steps
in computing the solution (\ref{ResultAns})--(\ref{ResultFerm2}).
We start with the conventional instanton solution and construct
the deformed instanton in perturbation theory in the
non(anti)commutativity parameter $C^{\a\b}$.

\noindent {\bf First order for the bosons:}

We begin by solving the equation for the first correction to the gauge
field $A_m^{(1)}$:
\bea {{(F^{(1)}_{\alpha\beta})}_{  a}}^{  b}&=&-
\frac{i}{2}C_{\alpha\beta}\delta^{  a}_{  b}\left( \frac{3\cdot
64\rho^4}{(x^2+\rho^2)^4} {\overline\xi}_\ald{\overline\xi}^\ald-
\frac{{\overline\chi}_i
\chi^i}{(x^2+\rho^2)^3} \right) \nonumber \\
{{(F^{(1)}_{\alpha\beta})}_{i}}^{j}&=&+
iC_{\alpha\beta}\frac{{\overline\chi}_i \chi^j}{(x^2+\rho^2)^3} \nonumber\\
{(F^{(1)}_{\alpha\beta})_{  a}}^{i}&=&
-C_{\alpha\beta}\frac{12\rho^2}{(x^2+\rho^2)^{7/2}}{\bar\xi}_{\dot
a}\chi^i \nonumber \\
{(F^{(1)}_{\alpha\beta})_{i}}^{  a}&=&
-C_{\alpha\beta}\frac{12\rho^2}{(x^2+\rho^2)^{7/2}} {\overline
\chi}_i{\overline\xi}^{\dot a} \, . \nonumber \eea
As discussed in Appendix \ref{AppB}, an ansatz
\bea A^{(1)}_m=C_{mn}\nabla_n \Phi^{(1)} \nonumber \eea
reduces the entire problem to solving a Poisson-type equation in
the instanton background, and the solution of the Poisson equation
is given by (\ref{ResultBos1}).

\noindent {\bf First order for the fermions:}

At the next step, we solve the equation for ${\bar\la}^{(1)}$:
\bea \sigma^m_{\alpha\ald}\nabla_m{\overline\la}^{(1)\ald}=
-\frac{i}{2}[A^{(1)}_m,\sigma^m_{\alpha\ald}{\overline\la}^{(0)\ald}]=
-\frac{i}{2}C_{mn}[(\nabla_n \Phi^{(1)}),
\sigma^m_{\alpha\ald}{\overline\la}^{(0)\ald}] \, .\nonumber \eea
Using the Fierz identity and equation of motion for ${\bar\la}^{(0)}$,
we reduce this equation to
\bea \sigma^m_{\alpha\ald}\nabla_m{\overline\la}^{(1)\ald}=
\frac{i}{2}{C_\alpha}^\beta\sigma^m_{\beta\ald}\nabla_m
\left[\Phi^{(1)},{\overline\la}^{(0)\ald}\right]. \nonumber \eea
With an ansatz \bea
{\overline\la}^{(1)\ald}={\overline\sigma}^{m\ald\alpha}{C_\alpha}^\beta
\nabla_m\Psi^{(1)}_\beta, \nonumber\eea %
it is reduced further to
\bea\label{PreLaplFerm1} \sigma^m_{\beta\ald}\nabla_m\Big[
-i{\overline\sigma}^{n\ald\gamma}\nabla_n\Psi^{(1)}_\gamma-\frac{1}{2}
\left[\Phi^{(1)},{\overline\la}^{(0)\ald}\right]\Big]=0. \eea
Notice that this derivation does not rely on the specific form of
${\overline\la}^{(0)}$ the steps above are essentially the same as
those described in more detail in section \ref{SectNSC} for the
special situation $\eta_\alpha=0$ .

Using an identity for sigma matrices:
$\sigma_m{\bar\sigma}_n=-\eta_{mn}+2\sigma^{mn}$ and
anti-self-duality of the undeformed solution (which leads to
$\sigma^{mn}\nabla_m\nabla_n=0$), we can express
(\ref{PreLaplFerm1}) as a Poisson equation:
\bea\label{LaplFerm1}
\nabla^2\Psi^{(1)}_\alpha=-\sigma^n_{\alpha\ald}\nabla_n
J^{(1)\ald}, \qquad
J^{(1)\ald}\equiv\frac{i}{2}\left[\Phi^{(1)},{\overline\la}^{(0)\ald}\right].
\nonumber \eea
Explicit evaluation of the current gives:
\bea
{(J^\ald)_a}^b&=&-\delta_a^b\frac{i{\overline\xi}^\ald{\overline
\chi}_i\chi^i}{(r^2+\rho^2)^3}; \hskip3cm
{(J^\ald)_i}^j=2\frac{i{\overline\xi}^\ald{\bar
\chi}_i\chi^j}{(r^2+\rho^2)^3} \nonumber \\
{(J_\ald)_a}^i
&=&-\eps_{\ald a}\frac{4\chi^i}{(r^2+\rho^2)^{7/2}}\Big[
(r^2+\rho^2)({\bar\zeta}_\ald{\overline\zeta}^\ald+\rho^2
\eta^\alpha\eta_\alpha)+4\rho^2{\overline\xi}_\bed{\overline\xi}^\bed
\Big]\\
{(J_\ald)_i}^a&=&- \, \delta^a_\ald \, \frac{4{\bar
\chi}_i}{(r^2+\rho^2)^{7/2}}\Big[
(r^2+\rho^2)({\overline\zeta}_\ald{\overline\zeta}^\ald+\rho^2
\eta^\alpha\eta_\alpha)+4\rho^2{\overline\xi}_\bed{\overline\xi}^\bed
\Big]. \nonumber \eea
We used the following relation:
$2x_m{\bar\zeta}{\bar\sigma}^m\eta={\bar\xi}_\ald{\bar\xi}^\ald-
{\bar\zeta}_\ald{\bar\zeta}^\ald-x^2\eta^\alpha\eta_\alpha$.
The solution of (\ref{LaplFerm1}) is given by (\ref{ResultFerm1}).

\noindent {\bf Second order for the bosons:}

Repeating the arguments of section \ref{SectNSC}, we find the
equation: \bea\label{SecOrdBosEqn} &&2(\nabla_{[m} A_{n]}^{(2)})^+
-\frac{i}{4}C_{kl}C_{kl}\Big[\nabla_{[m}(\Phi^{(1)}\nabla_{n]}
\Phi^{(1)})\Big]^+\nonumber\\
&&\qquad
+\frac{i}{2}C_{mn}C_{kl}\Big[\nabla_{[k}\Phi^{(1)}\nabla_{l]}
\Phi^{(1)}+
{\bar\sigma}^{k\ald\alpha}[{\overline\la}_\ald^{(0)},\nabla_l
\Psi^{(1)}_\alpha] \Big]=0 \, . \eea
In section \ref{SectNSC}, we put forward the following ansatz for
$A^{(2)}$:
\bea\label{A2OldAns}
A_m^{(2)}=\frac{i}{8}C_{kl}C_{kl}\Phi^{(1)}\nabla_m {\Phi}^{(1)}+
C_{mn}\nabla_n {\Phi}^{(2)}. \eea
With the ansatz, (\ref{SecOrdBosEqn}) was reduced to an Poisson
equation for ${\Phi}^{(2)}$. Notice that, in special situation the
superconformal mode were set to zero, the solution $\Phi^{(1)}$
was such that $\mbox{Tr}\Phi^{(1)}\nabla_m{\Phi}^{(1)}=0$. As
such, $A_m^{(2)}$ did not have a component proportional to the
identity matrix. In the presence of superconformal modes,
$\mbox{Tr}\Phi^{(1)}\nabla_m{\Phi}^{(1)}$ is no longer zero, and
we found it convenient to modify the ansatz (\ref{A2OldAns}) so
that the trace component in $A_m^{(2)}$ is avoided. The simplest
such modification is:
\bea\label{A2NewAns} A_m^{(2)}=
\frac{i}{16}C_{kl}C_{kl}\Phi^{(1)}{\stackrel{\leftrightarrow}{\nabla}}_m
{\Phi}^{(1)}+ C_{mn}\nabla_n {\Phi}^{(2)} \, , \eea
and it does not spoil the Poisson equation for ${\Psi}^{(2)}$, since the
difference between (\ref{A2NewAns}) and (\ref{A2OldAns}),
$
\delta A_m^{(2)}=-\frac{i}{16}C_{kl}C_{kl}
{\nabla}_m(\Phi^{(1)}{\Phi}^{(1)})
$
satisfies
$
(\nabla_{[m} \delta A_{n]}^{(2)})^+=0.
$
With the ansatz (\ref{A2NewAns}), we get the equation for
${\Phi}^{(2)}$:
\bea \nabla^2
{\Phi}^{(2)}=iC_{kl}\left\{\nabla_k\Phi^{(1)}\nabla_l\Phi^{(1)}+
{\overline\sigma}^{k\ald\alpha}[{\overline\la}^{(0)}_\ald,\nabla_l\Psi^{(1)}_\alpha]\right\}
\equiv J^{(2)} \, . \nonumber \eea
Explicit evaluation of the current $J^{(2)}$ yields:
\bea {(J^{(2)})_a}^b&=&
\frac{16iC_{mk}}{(\rho^2+r^2)^5}{({\overline\sigma}^{kn})^b}_a
x_mx_n {\overline\chi}_i\chi^i\Big[
\frac{{\bar\zeta}_\ald{\overline\zeta}^\ald}{\rho^2}(r^2+6\rho^2)+
(r^2-4\rho^2)\eta^\alpha\eta_\alpha-10\eta^\alpha
x_{\alpha\ald}{\overline\zeta}^\ald
\Big]\nonumber\\
&-&\frac{40i{\overline\chi}_i\chi^i}{(\rho^2+r^2)^4}\eps_{a\ald}\left[
{\overline\zeta}^\ald ({\overline x}C)^{b\alpha}\eta_\alpha+
{\overline\zeta}^b ({\bar x}C)^{\ald\alpha}\eta_\alpha\right]
\nonumber \\
{(J^{(2)})_i}^a&=&\frac{32{\overline
\chi}_i}{(r^2+\rho^2)^{9/2}}
\left[(r^2+9\rho^2){\overline\zeta}_\ald{\overline\zeta}^\ald
({\overline x}C)^{a\alpha}\eta_\alpha+
8\rho^2\eta^\alpha\eta_\alpha {({\overline x}C
x)^a}_\ald{\overline\zeta}^\ald
\right]\nonumber\\
{(J^{(2)})_a}^i&=&\frac{32{A}^i\eps_{a\bed}}{(r^2+\rho^2)^{9/2}}
\left[(r^2+9\rho^2){\bar\zeta}_\ald{\overline\zeta}^\ald ({\bar
x}C)^{\bed\alpha}\eta_\alpha+ 8\rho^2\eta^\alpha\eta_\alpha
{({\overline x}C x)^\bed}_\ald{\overline\zeta}^\ald
\right].\nonumber \eea
Solving the Poisson equation, we get (\ref{ResultBos2}).

\noindent {\bf Second order for the fermions:}

In the second order in $C$, we get the following equation for the
fermions:
\bea \sigma^m_{\alpha\ald}\nabla_m{\overline\la}^{(2)\ald}&=&-
\frac{i}{2}[A^{(1)}_m,\sigma^m_{\alpha\ald}{\overline\la}^{(1)\ald}]-
\frac{i}{2}[A^{(2)}_m,\sigma^m_{\alpha\ald}{\overline\la}^{(0)\ald}]\nonumber\\
&=& -\frac{i}{2}C_{mn}\nabla_n\left(
[\Phi^{(1)},\sigma^m_{\alpha\ald}{\overline\la}^{(1)\ald}]+
[\Phi^{(2)},\sigma^m_{\alpha\ald}{\overline\la}^{(0)\ald}]
\right)\nonumber\\
&&+\frac{i}{2}C_{mn}
[\Phi^{(1)},\sigma^m_{\alpha\ald}\nabla_n{\overline\la}^{(1)\ald}]
+\frac{1}{32}C_{kl}C_{kl}
[\Phi^{(1)}{\stackrel{\leftrightarrow}{\nabla}}_m
\Phi^{(1)},\sigma^m_{\alpha\ald}{\overline\la}^{(0)\ald}]  \, .
\nonumber \eea
After straightforward algebra, this equation is simplified as:
\bea \sigma^m_{\alpha\ald}\nabla_m{\overline\la}^{(2)\ald} =
-\frac{i}{2}C_{mn}\nabla_n\left(
[\Phi^{(1)},\sigma^m_{\alpha\ald}{\overline\la}^{(1)\ald}]+
[\Phi^{(2)},\sigma^m_{\alpha\ald}{\overline\la}^{(0)\ald}] \right)
-\frac{C_{kl}C^{kl}}{32}\sigma^m_{\alpha\ald}
\nabla_m[\Phi^{(1)},[\Phi^{(1)},{\overline\la}^{(0)\ald}]].
\nonumber \eea
This suggests the following ansatz:
\bea {\overline\la}^{(2)\ald}=-\frac{C_{kl}C^{kl}}{32}
[\Phi^{(1)},[\Phi^{(1)},{\overline\la}^{(0)\ald}]]+
{\overline\sigma}^{m\ald\alpha}{C_\alpha}^\beta\nabla_m\Psi^{(2)}_\beta,
\nonumber \eea
which leads to the following Poisson-type equation:
\bea -\nabla^2 \Psi^{(2)}_\alpha=\frac{i}{2}\sigma^m_{\alpha\ald}
\nabla_m\left([\Phi^{(1)},{\overline\la}^{(1)\ald}]+
[\Phi^{(2)},{\overline\la}^{(0)\ald}] \right). \nonumber \eea
The solution is given in (\ref{ResultFerm2}).

\noindent {\bf Third order for the bosons:}

Finally, in the third order in $C$, we need to solve for the gauge
fields only. The equations are
\bea 2(\nabla_{[m} A_{n]}^{(3)})^+
+\frac{i}{2}2[A^{(1)}_{[m},A^{(2)}_{n]}]^+ +\frac{i}{2}C_{mn}
({\overline\la}_\ald^{(2)}{\overline\la}^{(0)\ald}+
{\overline\la}_\ald^{(0)}{\overline\la}^{(2)\ald}+
{\overline\la}_\ald^{(1)}{\overline\la}^{(1)\ald})=0 \, .
\nonumber \eea
Direct computation of the current yields
\bea (\nabla_{m} A_{n}^{(3)}-\nabla_{n} A_{m}^{(3)})^+ &=&
-\frac{i}{2}C_{mn} \frac{C_{kl}C_{kl}}{4}
\frac{32\eta^\alpha\eta_\alpha{\bar\zeta}_\ald{\bar\zeta}^\ald{\bar
\chi}_i\chi^i}{
\rho^2(r^2+\rho^2)^5}\nonumber\\
&\times&
{\mbox{diag}}\Big(3(r^4-4r^2\rho^2-\rho^4),3(r^4-4r^2\rho^2-\rho^4),
2(r^4-10r^2\rho^2+\rho^4)\Big) \, , \nonumber \eea
where the nonzero entries reside in the U(3) block. This equation
is soluble by taking
\bea A^{(3)}_m=C_{mn}\d_n\Phi^{(3)} , \nonumber \eea
where
\bea \Phi^{(3)}=2i \frac{C_{kl}C_{kl}}{4}
\frac{\eta^\alpha\eta_\alpha{\overline\zeta}_\ald{\overline\zeta}^\ald{\overline
\chi}_i\chi^i}{ \rho^4(r^2+\rho^2)^3}\,
{\mbox{diag}}\Big(r^4+6r^2\rho^2+3\rho^4,r^4+6r^2\rho^2+3\rho^4,
2(r^4+4r^2\rho^2+\rho^4) \Big). \nonumber \eea


\begin{thebibliography}{999}

\bibitem{dijkgraafvafa}
R.~Dijkgraaf and C.~Vafa, {\sl Matrix models, topological strings,
and supersymmetric gauge theories},
Nucl.\ Phys.\ B {\bf 644}, 3 (2002) [arXiv:hep-th/0206255];\\
R.~Dijkgraaf and C.~Vafa, {\sl On geometry and matrix models},
Nucl.\ Phys.\ B {\bf 644}, 21 (2002) [arXiv:hep-th/0207106];\\
R.~Dijkgraaf and C.~Vafa, {\sl A perturbative window into
non-perturbative physics}, arXiv:hep-th/0208048.
R.~Dijkgraaf, M.~T.~Grisaru, C.~S.~Lam, C.~Vafa and D.~Zanon, {\sl
Perturbative computation of glueball superpotentials}, Phys.\
Lett.\ B {\bf 573}, 138 (2003) [arXiv:hep-th/0211017].

\bibitem{oogurivafa}
H.~Ooguri and C.~Vafa, {\sl The C-deformation of gluino and
non-planar diagrams}, Adv.\ Theor.\ Math.\ Phys.\  {\bf 7}, 53
(2003) [arXiv:hep-th/0302109];\\
H.~Ooguri and C.~Vafa, {\sl Gravity induced C-deformation},
arXiv:hep-th/0303063.

\bibitem{deboer}
J.~de Boer, P.~A.~Grassi and P.~van Nieuwenhuizen, {\em
Non-commutative superspace from string theory},
arXiv:hep-th/0302078.

\bibitem{seiberg}
N.~Seiberg, {\em Noncommutative superspace, N = 1/2 supersymmetry,
field theory and  string theory}, JHEP {\bf 0306}, 010 (2003)
[arXiv:hep-th/0305248].

\bibitem{berkovitsseiberg}
N.~Berkovits and N.~Seiberg, {\em Superstrings in graviphoton
background and N = 1/2 + 3/2 supersymmetry}, JHEP {\bf 0307}, 010
(2003) [arXiv:hep-th/0306226].

\bibitem{therest}
M.~Hatsuda, S.~Iso and H.~Umetsu, {\sl Noncommutative superspace,
supermatrix and lowest Landau level}, Nucl.\ Phys.\ B {\bf 671},
217 (2003)
[arXiv:hep-th/0306251];\\
T.~Araki, K.~Ito and A.~Ohtsuka, {\sl Supersymmetric gauge
theories on noncommutative superspace}, Phys.\ Lett.\ B {\bf 573},
209 (2003)
[arXiv:hep-th/0307076];\\
Y.~Shibusa and T.~Tada, {\sl Note on a fermionic solution of the
matrix model and noncommutative superspace},
arXiv:hep-th/0307236;\\
E.~Ivanov, O.~Lechtenfeld and B.~Zupnik, {\sl Nilpotent
deformations of N = 2 superspace},
arXiv:hep-th/0308012;\\
S.~Ferrara and E.~Sokatchev, {\sl Non-anticommutative N = 2
super-Yang-Mills theory with singlet deformation},
arXiv:hep-th/0308021;\\
R.~Abbaspur, {\sl Scalar solitons in non(anti)commutative
superspace}, arXiv:hep-th/0308050;\\
A.~Sako and T.~Suzuki, {\sl Ring structure of SUSY * product and
1/2 SUSY Wess-Zumino model},
arXiv:hep-th/0309076;\\
B.~Chandrasekhar and A.~Kumar, {\sl D = 2, N = 2, supersymmetric
theories on non(anti)commutative superspace},
arXiv:hep-th/0310137;\\
S.~Iso and H.~Umetsu, {\sl Gauge theory on noncommutative
supersphere from supermatrix model}, arXiv:hep-th/0311005;\\
A.~Imaanpur,
{\sl Comments on gluino condensates in N = 1/2 SYM theory},
arXiv:hep-th/0311137.


\bibitem{renorm}
R.~Britto, B.~Feng and S.~J.~Rey, {\sl Deformed superspace, N =
1/2 supersymmetry and (non)renormalization  theorems}, JHEP {\bf
0307}, 067 (2003) [arXiv:hep-th/0306215];\\
S.~Terashima and J.~T.~Yee, {\sl Comments on noncommutative
superspace},
arXiv:hep-th/0306237;\\
M.~T.~Grisaru, S.~Penati and A.~Romagnoni, {\sl Two-loop
renormalization for nonanticommutative N = 1/2 supersymmetric WZ
model}, JHEP {\bf 0308}, 003 (2003)
[arXiv:hep-th/0307099];\\
R.~Britto and B.~Feng, {\sl N = 1/2 Wess-Zumino model is
renormalizable}, Phys.\ Rev.\ Lett.\  {\bf 91}, 201601 (2003)
[arXiv:hep-th/0307165];\\
A.~Romagnoni, {\sl Renormalizability of N = 1/2 Wess-Zumino model
in superspace}, JHEP {\bf 0310}, 016 (2003)
[arXiv:hep-th/0307209];\\
O.~Lunin and S.~J.~Rey, {\sl Renormalizability of
non(anti)commutative gauge theories with N = 1/2 supersymmetry},
JHEP {\bf 0309}, 045 (2003)
[arXiv:hep-th/0307275];\\
M.~Alishahiha, A.~Ghodsi and N.~Sadooghi, {\sl One-loop
perturbative corrections to non(anti)commutativity parameter of N
= 1/2 supersymmetric U(N) gauge theory}, arXiv:hep-th/0309037.

\bibitem{berey}
D.~Berenstein and S.~J.~Rey, {\sl Wilsonian proof for
renormalizability of N = 1/2 supersymmetric field theories}, to be
published in Phys. Rev. {\bf D} (2003) [arXiv:hep-th/0308049].


\bibitem{Hitchin}
N.J. Hitchin, {\sc The Geometry and Topology of Moduli Spaces}, in
{\sc Global Geometry and Mathematical Physics}, Lecture Notes in
Mathematics 1451 (Springer, Heidelberg, 1988), 1-48;
%
%

\bibitem{ADHM}
M.~F.~Atiyah, N.~J.~Hitchin, V.~G.~Drinfeld and Y.~I.~Manin,
{\sl Construction Of Instantons},
Phys.\ Lett.\ A {\bf 65}, 185 (1978).

\bibitem{imaanpur}
A.~Imaanpur, {\sl On instantons and zero modes of N = 1/2 SYM
theory}, JHEP {\bf 0309}, 077 (2003) [arXiv:hep-th/0308171].

\bibitem{grassi}
P.~A.~Grassi, R.~Ricci and D.~Robles-Llana, {\sl Instanton
calculations for N = 1/2 super Yang-Mills theory},
arXiv:hep-th/0311155.

\bibitem{polchinski}
J.~Polchinski, {\sl Scale And Conformal Invariance In Quantum
Field Theory}, Nucl.\ Phys.\ B {\bf 303}, 226 (1988).

\bibitem{maldacena}
J.~M.~Maldacena, {\sl The large N limit of superconformal field
theories and supergravity}, Adv.\ Theor.\ Math.\ Phys.\  {\bf 2},
231 (1998) [Int.\ J.\ Theor.\ Phys.\  {\bf 38}, 1113 (1999)]
[arXiv:hep-th/9711200].

\bibitem{GKPW}
S.~S.~Gubser, I.~R.~Klebanov and A.~M.~Polyakov,
{\sl Gauge theory correlators from non-critical string theory},
Phys.\ Lett.\ B {\bf 428}, 105 (1998)
[arXiv:hep-th/9802109];\\
E.~Witten,
{\sl Anti-de Sitter space and holography},
Adv.\ Theor.\ Math.\ Phys.\  {\bf 2}, 253 (1998)
[arXiv:hep-th/9802150].
%
%
\bibitem{balasubramanian}
S.-J. Rey, unpublished work (1998);\\
V.~Balasubramanian, P.~Kraus, A.~E.~Lawrence and S.~P.~Trivedi,
{\sl Holographic probes of anti-de Sitter space-times},
Phys.\ Rev.\ D {\bf 59}, 104021 (1999) [arXiv:hep-th/9808017];\\
see also \cite{narain}.
%
\bibitem{MinwSeib}
S.~Minwalla, M.~Van Raamsdonk and N.~Seiberg,
{\sl Noncommutative perturbative dynamics},
JHEP {\bf 0002}, 020 (2000)
[arXiv:hep-th/9912072].
%
\bibitem{owls}
S.~R.~Das and S.~J.~Rey, {\sl Open Wilson lines in noncommutative
gauge theory and tomography of  holographic dual supergravity},
Nucl.\ Phys.\ B {\bf 590}, 453 (2000) [arXiv:hep-th/0008042];\\
D.~J.~Gross, A.~Hashimoto and N.~Itzhaki, {\sl Observables of
non-commutative gauge theories}, Adv.\ Theor.\ Math.\ Phys.\ {\bf
4}, 893 (2000) [arXiv:hep-th/0008075];\\
Y.~Kiem et.al., {\sl Open Wilson lines and generalized star
product in nocommutative scalar  field theories}, Phys.\ Rev.\ D
{\bf 65}, 026002 (2002)
[arXiv:hep-th/0106121];\\
ibid., {\sl Anatomy of one-loop effective action in noncommutative
scalar field  theories}, Eur.\ Phys.\ J.\ C {\bf
22}, 757 (2002) [arXiv:hep-th/0107106];\\
ibid., {\sl Anatomy of two-loop effective action in noncommutative
field theories}, Nucl.\ Phys.\ B {\bf 641}, 256 (2002)
[arXiv:hep-th/0110066];\\
ibid., {\sl Interacting open Wilson lines in noncommutative field
theories}, Phys.\ Rev.\ D {\bf 65}, 046003 (2002) [arXiv:hep-th/0110215];\\
ibid., {\sl Open Wilson Lines As Closed Strings In Non-Commutative
Field Theories}, Eur.\ Phys.\ J.\ C {\bf 22}, 781 (2002).

\bibitem{leshouchesreview}
S.~J.~Rey, {\sl Exact answers to approximate questions:
Noncommutative dipoles, open  Wilson lines, and UV-IR duality},
arXiv:hep-th/0207108;\\
ibid., J.\ Korean Phys.\ Soc.\  {\bf 39}, S527 (2001).
%
%
\bibitem{brittofengrey2}
R.~Britto, B.~Feng and S.~J.~Rey, {\sl Non(anti)commutative
superspace, UV/IR mixing and open Wilson lines}, JHEP {\bf 0308},
001 (2003) [arXiv:hep-th/0307091].

\bibitem{InformMetr}
M.K. Murray, {\sl The information metric on rational maps}, Exp.
Math. {\bf 2}, 271 (1994);\\
D. Groisser and M.K. Murray, {\sl Instantons and the information
metric}, [arXiv:dg-ga/9611008];\\
S.~Parvizi, {\em Non-commutative instantons and the information
metric}, Mod.\ Phys.\ Lett.\ A {\bf 17}, 341 (2002)
[arXiv:hep-th/0202025].

\bibitem{narain} M.~Blau, K.~S.~Narain and G.~Thompson,
{\em Instantons, the information metric, and the AdS/CFT
correspondence}, [arXiv:hep-th/0108122].

\bibitem{shifman}
M.~A.~Shifman and A.~I.~Vainshtein,
{\sl Instantons versus supersymmetry: Fifteen years later},
arXiv:hep-th/9902018.

\end{thebibliography}
\end{document}